\documentclass{emulateapj}

\usepackage{amsmath}
\usepackage{epsfig}
\usepackage{amssymb}
\usepackage{graphicx}
\usepackage{color}
\usepackage{natbib}

\def\gtaprx {\lower .1ex\hbox{\rlap{\raise .6ex\hbox{\hskip .3ex
	{\ifmmode{\scriptscriptstyle >}\else
		{$\scriptscriptstyle >$}\fi}}}
	\kern -.4ex{\ifmmode{\scriptscriptstyle \sim}\else
		{$\scriptscriptstyle\sim$}\fi}}}
\def\ltaprx {\lower .1ex\hbox{\rlap{\raise .6ex\hbox{\hskip .3ex
	{\ifmmode{\scriptscriptstyle <}\else
		{$\scriptscriptstyle <$}\fi}}}
	\kern -.4ex{\ifmmode{\scriptscriptstyle \sim}\else
		{$\scriptscriptstyle\sim$}\fi}}}

\newcommand{\cutt}[1]{\textcolor{blue}{}}

\newcommand{\Ms}{$M_{\odot}$}
\newcommand{\Zs}{$Z_{\odot}$}

\newcommand{\Ni}{{\ensuremath{^{56}\mathrm{Ni}}}}

\begin{document}

\title{Finding the First Cosmic Explosions I: Pair-Instability Supernovae}

\author{Daniel J. Whalen\altaffilmark{1,2}, Wesley Even\altaffilmark{3}, Lucille H. 
Frey\altaffilmark{4,5}, Joseph Smidt\altaffilmark{1}, Jarrett L. Johnson\altaffilmark{6}, C. C. 
Lovekin\altaffilmark{1}, Chris L. Fryer\altaffilmark{3}, Massimo Stiavelli\altaffilmark{7}, Daniel 
E. Holz\altaffilmark{8}, Alexander Heger\altaffilmark{9}, S. E. Woosley\altaffilmark{10} and 
Aimee L. Hungerford\altaffilmark{6}}

\altaffiltext{1}{T-2, Los Alamos National Laboratory, Los Alamos, NM 87545}

\altaffiltext{2}{Universit\"{a}t Heidelberg, Zentrum f\"{u}r Astronomie, Institut f\"{u}r 
Theoretische Astrophysik, Albert-Ueberle-Str. 2, 69120 Heidelberg, Germany}

\altaffiltext{3}{CCS-2, Los Alamos National Laboratory, Los Alamos, NM 87545}

\altaffiltext{4}{HPC-3, Los Alamos National Laboratory, Los Alamos, NM 87545}

\altaffiltext{5}{Department of Computer Science, University of New Mexico, Albuquerque, NM  
87131}

\altaffiltext{6}{XTD-6, Los Alamos National Laboratory, Los Alamos, NM 87545}

\altaffiltext{7}{Space Telescope Science Institute, 3700 San Martin Drive, Baltimore, MD 21218}

\altaffiltext{8}{Enrico Fermi Institute, Department of Physics, and Kavli Institute for Cosmological 
Physics, University of Chicago, Chicago, IL 60637, USA}

\altaffiltext{9}{Monash Centre for Astrophysics, Monash University, Victoria, 3800, Australia}

\altaffiltext{10}{Department of Astronomy and Astrophysics, UCSC, Santa Cruz, CA  95064}

\begin{abstract}

The first stars are the key to the formation of primitive galaxies, early cosmological reionization 
and chemical enrichment, and the origin of supermassive black holes.  Unfortunately, in spite 
of their extreme luminosities, individual Population III stars will likely remain beyond the reach 
of direct observation for decades to come. However, their properties could be revealed by their 
supernova explosions, which may soon be detected by a new generation of NIR observatories 
such as \textit{JWST} and WFIRST.  We present light curves and spectra for Pop III pair-instability 
supernovae calculated with the Los Alamos radiation hydrodynamics code RAGE.  Our numerical 
simulations account for the interaction of the blast with realistic circumstellar envelopes, the opacity 
of the envelope, and Lyman absorption by the neutral IGM at high redshift, all of which are crucial 
to computing the NIR signatures of the first cosmic explosions.  We find that \textit{JWST} will 
detect pair-instability supernovae out to $z \ga$ 30, WFIRST will detect them in all-sky surveys out 
to $z \sim$ 15 - 20 and LSST and Pan-STARRS will find them at $z \la 7 - 8$.  The discovery of 
these ancient explosions will probe the first stellar populations and reveal the existence of primitive 
galaxies that might not otherwise have been detected. 

\vspace{0.1in}

\end{abstract}


\section{Introduction}

Population III (Pop III) stars are the key to understanding primeval galaxies \citep{jgb08,
get08,jlj09,get10,jeon11,pmb11,pmb12,wise12}, the chemical enrichment and reionization 
of the early IGM \citep{ss07,bsmith09,chiaki12,ritt12,ss13}, and the origin of supermassive 
black holes \citep{bl03,jb07b,brmvol08,milos09,awa09,lfh09,th09,pm11,pm12,jlj12a,wf12,
agarw12,jet13,pm13,latif13c,latif13a,schl13,choi13}.  Unfortunately, even though they are 
thought to be extremely luminous \citep{s02}, individual Pop III stars will not be visible to 
the \textit{James Webb Space Telescope} \citep[\textit{JWST},][]{jwst06}, the Wide-Field 
Infrared Survey Telescope (WFIRST) or the Thirty-Meter Telescope (TMT), \citep[but 
see][on the possibility of detecting Pop III star H II regions by strong gravitational lensing]{
rz12}.  

Numerical simulations suggest that the first stars are born in 10$^5$ - 10$^6$  \Ms\ dark 
matter halos at $z \sim$ 20 - 30.  The original models implied that they are 100 - 500 
\Ms\ and form in isolation, one per halo \citep{bcl99,abn00,abn02,bcl02,nu01,on07,y08}, 
but newer models have since shown that some Pop III stars form in binaries \citep{turk09} 
and perhaps even in small clusters \citep{stacy10,clark11,get11,get12,susa13}.  
Simulations of UV breakout from primordial star-forming disks find that radiative feedback 
limits the final masses of some Pop III stars to $\lesssim$ 40 \Ms \citep{hos11,stacy12,
hos12} \citep[but also see][]{op01,oi02,op03,tm04,tm08,hir13}.  However, none of these 
models realistically bridges the gap between the formation and fragmentation of the 
protostellar disk and its evaporation up to a Myr later, and they rely on uniform accretion 
rates and simple recipes for protostellar evolution.  For these reasons, and because the 
roles of turbulence \citep{latif13a}, magnetic fields \citep{schob12} and radiation transport 
in the formation and evolution of primordial disks is not understood, numerical models 
cannot yet constrain the Pop III initial mass function (IMF) \citep[for recent reviews, see][]{
dw12,glov12}.

Some have tried to infer the masses of Pop III stars from their nucleosynthetic imprint on 
later generations, some of which may live today as dim metal-poor stars in the Galactic 
halo \citep[e.g.,][]{bc05,fet05,caffau12}.  The current consensus is that 15 - 40 \Ms\ Pop III 
stars die in core-collapse supernovae (CC SNe) and 140 - 260 \Ms\ stars explode as far 
more energetic pair-instability (PI) SNe, with up to 100 times the energy of Type Ia or Type 
II explosions \citep{hw02}. \citet{jet09b} (hereafter JET10) recently found that the yields of 
15 - 40 \Ms\ Pop III CC SNe are consistent with the chemical abundances measured in a 
sample of $\sim$ 130 extremely metal-poor stars \citep{Cayrel2004,Lai2008}.  But traces 
of the distinctive "odd-even" nucleosynthetic fingerprint of Pop III PI SNe have now been 
found in high-redshift damped Lyman alpha absorbers \citep{cooke11}.  Eighteen low
metallicity stars recently discovered in the \textit{Sloan Digital Sky Survey} have also been 
designated for further spectroscopic followup because they too are suspected to exhibit this 
pattern \citep{ren12} \citep[see also][on why the odd-even effect may not have been found 
in earlier surveys]{karl08}.  This new evidence from the fossil abundance record indicates 
that both low-mass and very massive Pop III stars existed in the early universe.  Several
stars above the classical 150 \Ms\ limit have also now been discovered at metallicities $Z 
\sim$ 0.1 \Zs\ in the R136 cluster, including one 300 \Ms\ candidate, further corroborating 
the possibility of very massive star formation \citep{R136}.

Detections of Pop III SNe will be the most direct probe of the first stars in the near term 
because they are thousands of times brighter than their progenitors and the primitive 
galaxies that host them \citep{byh03,ky05,get07,wet08a,ds11a,vas12,pan12b}.  PI SNe in 
particular are ideal candidates for finding Pop III stars because of their extreme luminosities.  
Besides the newest results from the fossil abundance record, other new discoveries suggest 
that PI SNe are more frequent at high redshifts than previously thought.  It is now known that
rotating Pop III stars can die as PI SNe at masses as low as 85 \Ms\ \citep{yoon06,cw12,
yoon12}.  Assuming simple power-law IMFs, this could increase PI SN rates in the early 
universe by a factor of 4 \citep[see also][on the effects of rotation and magnetic fields on 
Pop III star evolution]{Ekstr08,stacy11b,stacy13}.  Perhaps the most compelling evidence
is that a PI SN candidate has now been discovered in the local universe \citep[SN 2007bi,][]{
gy09,yn10}, in environments that are far less favorable to the formation of massive 
progenitors than at early epochs \citep[although see][for alternative interpretations of this
event]{kbi10,det12a}.

Previous studies have addressed detection thresholds for PI SNe at $z \sim$ 5 \citep{
sc05}, $6 < z < 15$ \citep{pan12a}, and $z \sim 30$ \citep[in approximate terms;][]{hum12} 
\citep[see also][]{kasen11,det12,ds13}.  The interaction of the PI SN with its envelope, the 
opacity of the envelope, and Lyman absorption by the neutral IGM at early epochs must all 
be taken into account to calculate its near infrared (NIR) signature at high redshift.  Doing 
so, \citet{wet12a} found that \textit{JWST} will detect PI SNe out to $z \sim$ 30.  However, 
their models did not cover the full range of stellar structures expected for 140 - 260 \Ms\ 
Pop III stars.  We have now calculated Pop III PI SNe for both blue giants and red 
hypergiants at $7 < z < 30$ with the Los Alamos RAGE and SPECTRUM codes.  In Section 
2 we review the PI explosion mechanism, the presupernova structures of the stars, and our 
Kepler Pop III PI progenitor and SN models.  We describe our RAGE and SPECTRUM 
source frame light curve and spectrum calculations in Section 3, and we examine PI SN 
blast profiles and spectra in Section 4.  In Section 5 we present Pop III PI SN NIR light 
curves and determine their detection limits as a function of redshift.  In Section 6 we discuss 
PI SN detection rates at $5 < z < 30$, and conclude in Section 7.

\begin{figure*}
\plottwo{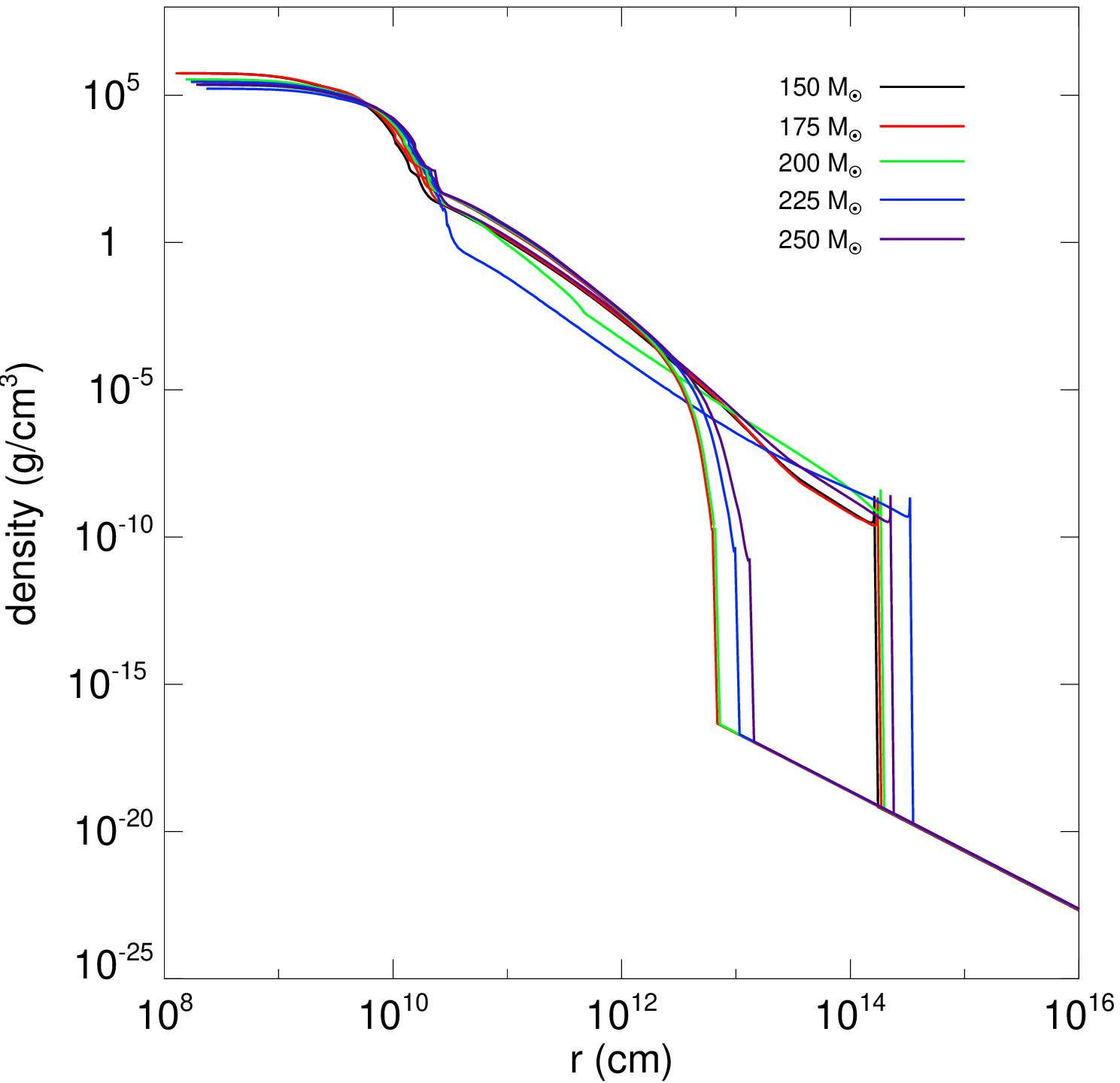}{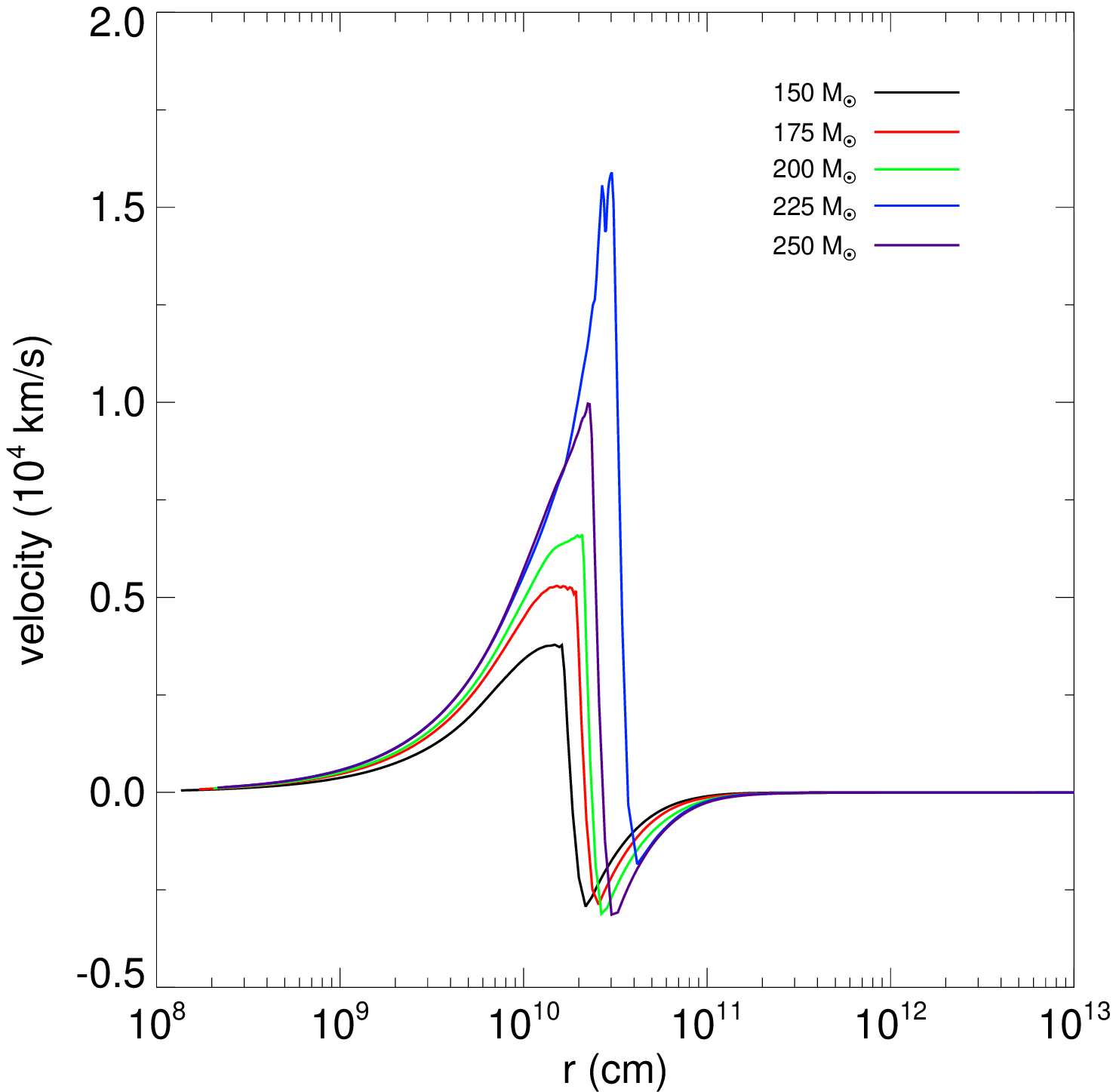}
\caption{Kepler PI SN explosion and stellar envelope profiles. Left panel: densities for all nine 
models.  The surface of each star is marked by the abrupt drop in density by ten orders of 
magnitude at $r \sim$ 10$^{13}$ cm in the four z-series profiles and a few $\times$ 10$^{14}$ 
cm in the five u-series stars.  A wind with a total mass of 0.1 \Ms, a velocity of 10$^8$ 
cm/s, and a free-streaming $r^{-2}$ profile is extended from the surface of the star out to the 
uniform relic H II region surrounding the star, which is not visible on this plot.  Right panel:  
velocity profiles for the u-series stars. Both the outgoing shock and ongoing collapse of the 
outer layers of the star are visible.  
\vspace{0.1in}}
\label{fig:PISNprof}
\end{figure*}

\section{PI SN Models}

PI SNe were first proposed by \citet{rs67} and \citet{brk67} and have been studied numerous 
times since then \citep[see][and references therein]{hw02}.  Pop III stars above 65 \Ms\ encounter 
the pair instability after central carbon burning, when thermal energy creates $e^{+} e^{-}$ pairs 
rather than maintaining pressure support against collapse. The cores of these stars subsequently 
contract, triggering explosive thermonuclear burning of O and Si.  Above 140 \Ms, the energy that
is released completely unbinds the star, and no black hole is formed.  At 260 \Ms\ the core of the 
star reaches temperatures that are high enough for alpha particles to be photo-disintegrated into 
free nucleons, which consumes as much energy per unit mass as was released by all preceding 
burning.  The star collapses instead of exploding.  PI SNe are the most energetic thermonuclear 
events in the universe, with yields of up to 10$^{53}$ erg for 260 \Ms\ stars.

The early spectral signatures of Pop III PI SNe heavily depend on the radius and structure of the 
star in addition to the explosion energy, the interaction of the blast with the envelope, the opacity 
of the envelope, and absorption by the IGM.  Shock temperatures at breakout are lower for large 
stars than for compact stars of equal mass because the shock has expanded to a greater radius 
and done more work on its environment.  Early spectra from the explosions of compact blue stars
are therefore harder than those of red giants.  As we discuss below, the structure of the star also 
determines the emission lines that appear in its spectra over time.  The size and structure of the 
progenitor in turn are governed by its metallicity and by internal convection over its life.

\subsection{Semi-Convective Mixing}

Convection can determine if a very massive Pop III star dies as a compact blue giant or a red 
hypergiant.  As described in detail in \citet{sc05}, the convection zone of the star can approach, 
touch or even penetrate the lower hydrogen layers, mixing them with carbon dredged up from 
the core from He burning.  When these two high temperature components mix, they burn 
violently, boosting energy production rates in the H shell by up to several orders of magnitude.  
This, together with the now greater opacity of the lower hydrogen layer, can puff up the star by 
more than an order of magnitudein radius.   

\subsection{Metallicity}

Gas in high-redshift halos that is enriched to metallicities below 10$^{-3.5}$ \Zs\ fragments 
on mass scales that are essentially identical to those of pristine gas and forms very massive 
stars \citep[e.g.][]{bcl01,mbh03,ss07}. However, these low metallicities are more than enough 
to enhance CNO reaction and energy production rates in the hydrogen burning layers of the 
star, inflating the stellar envelope as much as convection.  Since there is a strong degeneracy 
between the effects of metals and convection on the structure of the star, the full range of light 
curves and spectra for Pop III PI SNe is as easily spanned by metallicity as by convective 
overshoot, as we demonstrate below. 

\subsection{Explosive Mixing}

JET10 found that in 15 - 40 \Ms\ Pop III SNe the shock completely disrupts the interior of the 
star, heavily mixing the ejecta by the time it ruptures the surface.  In contrast, the concentric
shells of elements expand almost homologously in PI SNe, with only occasional minor mixing 
between the O and He layers prior to breakout \citep{jw11} \citep[see also][]{chen11}.  Mixing 
in the star during the explosion can determine the order in which elements appear in emission 
lines over time.  In the frame of the shock, the photosphere from which photons escape 
descends into the ejecta over time as each fluid element expands and the ejecta is diluted.  If 
Rayleigh-Taylor (RT) instabilities drive mixing prior to shock breakout, they can dredge heavy 
elements from a deeper mass coordinate up to the photosphere and expose them to the IGM 
at much earlier times.  Since mixing is minimal in Pop III PI SNe, the absence of metal lines 
soon after breakout would be one of several markers of the event.  

\subsection{Kepler}

To model the structure of the progenitor, we evolve 150, 175, 200, 225, and 250 \Ms\ 
zero-metallicity stars (z-series) and 10$^{-4}$ \Zs\ stars (u-series) from the zero age main 
sequence to the onset of collapse in the one-dimensional (1D) Lagrangian stellar evolution 
code Kepler \citep{Weaver1978,Woosley2002}.  The explosion begins when this collapse 
triggers rapid O and Si burning.  Unlike the CC SN simulations of JET10, in which the blast 
must be artificially launched with a piston and the explosion energy is a free parameter, the 
PI SN is an emergent feature of our stellar evolution model, and its energy is set by how 
much O and Si burns.  The blast was followed until the end of all nuclear burning at $\sim$ 
20 s, when the shock was still deep inside the star.  We calculate energy generation with a 
19-isotope network up to the point of oxygen depletion in the core and with a 128-isotope 
quasi-equilibrium network thereafter.  The z-series 150 \Ms\ star collapses to a black hole 
without an explosion. The number of mass zones on the grid ranged from 1000 - 1200 and 
was always sufficient to resolve all salient structures of the star and SN.  We show density 
and velocity profiles for our explosions in Fig.~\ref{fig:PISNprof} and summarize our grid 
of models in Table \ref{tab:table1}.  We consider only non-rotating progenitors.

We use metallicity rather than convective overshoot to parametrize the progenitors because 
for a given star the size of the central convection zone is uncertain but $Z$ = 0 and 10$^{-4}$ 
\Zs\ bracket the metallicities over which 140 - 260 \Ms\ Pop III stars are expected to form 
\citep{mbh03,ss07,bsmith09}.  As we show in Fig.~\ref{fig:PISNprof}, these two metallicities 
yield blue giant and red hypergiant envelopes similar to those obtained by varying convective 
mixing in \citet{sc05}.  All u-series stars in our study die as red hypergiants and all z-series 
stars die as blue giants.  The light curves and spectra for our SNe should therefore bracket 
those that will be observed (but most 140 - 260 \Ms\ Pop III stars are thought to have
convective mixing and die as red stars rather than blue stars).

\begin{deluxetable}{lcccc}  
\tabletypesize{\scriptsize}  
\tablecaption{Kepler PI SN Models (masses are in \Ms\label{tab:table1})}
\tablehead{
\colhead{model} & \colhead{$M_{He}$}& \colhead{$R$ ($10^{13}$ cm)}& \colhead{$E$ 
(10$^{51}$ erg)} & \colhead{$M_{\Ni}$}}
\startdata 
u150 &  72      &  16.2   &   9.0    &  0.07   \\
u175 &  84.4   &  17.4   &   21.3  &  0.70   \\
u200 &  96.7   &  18.4   &   33     &  5.09   \\
u225 &  103.5 &  33.3   &   46.7  &  16.5   \\
u250 &  124    &  22.5   &   69.2  &  37.9   \\
z175 &  84.3   &   0.62  &   14.6  &  0        \\
z200 &  96.9   &   0.66  &   27.8  &  1.9     \\
z225 &  110.1 &   0.98  &   42.5  &  8.73   \\
z250 &  123.5 &   1.31  &   63.2  &  23.1      
\enddata 

\end{deluxetable}  
\vspace{0.3in}

\section{RAGE and SPECTRUM Simulations}

We propagate the shock through the interior of the star, its surface, and then out into the 
surrounding medium with the radiation hydrodynamics code RAGE \citep{rage}.  RAGE 
(Radiation Adaptive Grid Eulerian) is a multidimensional adaptive mesh refinement (AMR) 
radiation hydrodynamics code developed at Los Alamos National Laboratory (LANL).  
RAGE couples second-order conservative Godunov hydrodynamics to grey or multigroup 
flux-limited diffusion (FLD) to model strongly radiating flows.  RAGE utilizes the extensive 
LANL OPLIB database of atomic 
opacities\footnote{http://aphysics2/www.t4.lanl.gov/cgi-bin/opacity/tops.pl} \citep{oplib} 
and can also evolve multimaterial flows with a variety of equations of state (EOS).  We 
describe most of the physics implemented in our RAGE models and why it is needed to 
capture the features of our light curves in detail in \citet{fet12} (hereafter FET12): 
multispecies advection, grey FLD radiation transport with 2-temperature (2T) physics, and 
energy deposition by radioactive decay of \Ni{}.  In particular, 2T radiation transport, in 
which radiation and matter temperatures are evolved separately, better models shock 
breakout and its aftermath when matter and radiation can be out of equilibrium.  This is an 
important improvement over earlier 1T models of PI SN explosions.  We evolve mass 
fractions for 15 elements, the even numbered elements predominantly synthesized in PI 
SNe.  

\subsection{Self-Gravity}

We have also now implemented self-gravity in RAGE, which was not included in the \citet{
wet12a} models.  Although they are extremely energetic, PI SN shocks are launched from 
deep inside the star where the gravitational potential energy of the ejecta is quite large.  If 
this energy is not taken into account the shock can break out of the star with too large a
velocity and luminosity.  Gravity was implemented in spherical symmetry by computing the 
potential
\begin{equation}
\phi \; = \; - \frac{GM_{\mathrm{encl}}}{r},
\end{equation}
where 
\begin{equation}
M_{\mathrm{encl}} \; = \; \int_0^r 4 \pi {r'}^2 \rho(r') dr'
\end{equation}
is constructed by extracting the densities from the finest levels of the AMR hierarchy and
reordering them by radius.  The gravitational potential is then applied to updates to the 
gas velocities and total energies every time step.  

\begin{figure}
\plotone{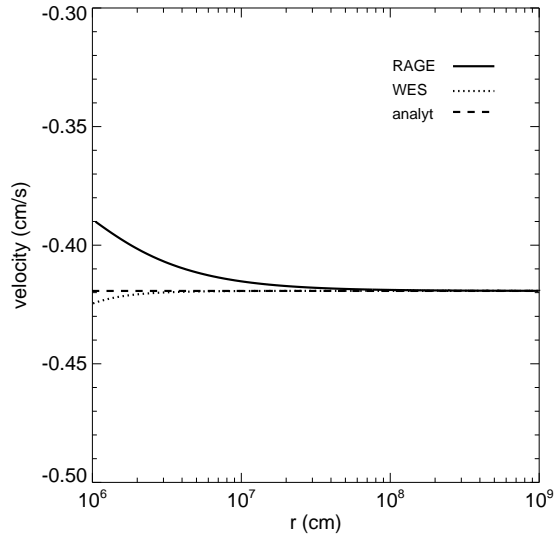}
\caption{Collapse test for the pressureless sphere, with $\alpha =$ 10$^4$.  Velocities 
are shown as a function of radius at t = 100 s ($v_{\mathrm{an}} = -0.419$ cm s$^{-1}$)
for RAGE and a semi-analytic code versus the analytical solution.
\vspace{0.1in}}
\label{fig:selfgrav}
\end{figure}

We tested gravity by running a pressureless sphere collapse problem with an analytical
solution.  If the sphere has the density profile
\begin{equation}
\rho(r) \; = \; \frac{\alpha}{r}
\end{equation}
then $M_{\mathrm{encl}} \; = \; 2 \pi \alpha r^2$ and in the absence of pressure forces 
each spherical shell experiences a constant acceleration $a$ given by
\begin{equation}
a \; = \; - \frac{GM_{\mathrm{encl}}}{r^2} \; = \; -2 \pi \alpha G.
\end{equation}
The sphere collapses homologously with a velocity $v \; = \; -2 \pi \alpha G t$.  We show
a snapshot of radial velocities for the collapse of the sphere for $\alpha =$ 10$^4$ at 100
s in Fig. \ref{fig:selfgrav}.  The sphere is zoned into 10000 uniform mesh points with 
reflecting inner boundary conditions, an outer boundary at 10$^9$ cm, a resolution of 10$
^5$ cm, and $v_{\mathrm{init}} =$ 0.  Although it is not possible to disable pressure forces 
in RAGE, we make them negligible by reducing the heat capacity $C_{\mathrm{V}}$ from 
the usual 1.2472 $\times$ 10$^8$ erg K$^{-1}$ for an ideal gas to 1.0 erg K$^{-1}$.

As shown in Fig. \ref{fig:selfgrav}, RAGE agrees with the analytical solution to within 1\% 
beyond 10$^7$ cm, with at most 10\% error at the inner boundary where gas begins to
pile up.  The departures from the analytic solution there are due to residual pressure 
forces that build up in the increasingly dense gas.  In this test problem RAGE conserves
total energy (gravitational $+$ kinetic $+$ internal) to within machine precision.

\subsection{Initial Grid}

We map densities, velocities, specific internal energies (erg gm$^{-1}$), and species mass 
fractions for the explosion and star from Kepler together with a circumstellar envelope onto 
a uniform 50,000 zone 1D spherical mesh in RAGE.  Since radiation energy densities are 
not explicitly evolved in the Kepler models, we initialize them in RAGE as 
\vspace{0.05in}
\begin{equation}
e_{rad} = aT^4, \vspace{0.05in}
\end{equation}
where $a =$ 7.564 $\times$ 10$^{-15}$ erg cm$^{-3}$ K$^{-4}$ is the radiation constant and 
$T$ is the gas temperature.  Also, because the gas energy in Kepler includes contributions by 
ionization states of atoms, we unambiguously construct the specific internal energy from $T$ 
with
\vspace{0.05in}
\begin{equation}
e_{gas} = C_VT, \vspace{0.05in}
\end{equation}
where $C_V = $ 1.2472 $\times$ 10$^{8}$ erg K$^{-1}$ is the specific heat of the gas.  Since 
there is little mixing in the star as the shock propagates to the surface, the radial distribution 
of elements in the star is essentially frozen in mass coordinate at death and expands 
homologously into space thereafter. 1D simulations are therefore sufficient to capture the key 
attributes of PI SN light curves and spectra. 

At the beginning of the simulation we allocate 5000 zones from the center of the grid to the 
edge of the shock in the velocity profile.  We allow up to five levels of refinement in the initial 
interpolation of the profile onto this grid and then throughout the simulation.  Our choice of 
grid ensures that the photosphere of the ejecta is always fully resolved, since failure to do so 
can lead to luminosity underestimates during post processing.  We impose reflecting and 
outflow boundary conditions on the fluid and radiation flows at the inner and outer boundaries 
of the mesh, respectively.  

When the calculation is begun, Courant times are initially small due to high temperatures, 
large velocities and small cell sizes.  To reduce execution times and to accommodate the 
expansion of the ejecta, we periodically regrid the profiles onto a larger mesh as described 
in detail in FET12.  Each time we regrid the blast we allocate 5000 zones out to either the 
edge of the shock (pre-breakout) or to the edge of the radiation front (post-breakout -- we 
take the the edge of the front to be where the radiation temperature falls to a few tenths of 
an eV.  Up to five levels of refinement are used during the regrid and then again during the
simulation.  The inner boundary is always at 0 cm and the outer boundary of the final, 
largest mesh in our models is 1.0 $\times$ 10$^{18}$ cm. We assume that the final mass 
of each star is the same as its initial mass.

\subsection{Circumstellar Winds}

Pop III stars are usually thought to die in low-density H II regions with relatively flat density 
profiles $n \sim$ 0.1 - 1 cm$^{-3}$ out to 100 - 200 pc \citep{wan04,ket04,abs06,awb07,
wa08a,wn08a,wn08b}.  They are not believed to lose much mass over their lifetimes 
because there are no metals in their atmospheres to drive strong winds \citep{Kudritzki00,
Baraffe01,Vink01,kk06,Ekstr08}.  However, we allow for the possibility that some Pop III
stars had winds.  First, mass loss from very massive stars is usually just parametrized by 
metallicity, not calculated from first principles with radiation hydrodynamics \citep[e.g.,][]{
mm94}.  Extrapolating no winds at zero metallicity could exclude mass loss by other means, 
such as helium opacity, hydromagnetic flows, or pulsational ejections \citep{hw02}. Second, 
mixing in rotating Pop III stars, which we do not consider here, can dredge metals up from 
the interior of the star to its surface and drive winds later in its life \citep[e.g.,][]{Ekstr08}, 
although the mass loss so far has been found to be minor. Lastly, there are no observations 
to rule out winds from very-massive zero-metallicity stars.  We therefore extend a low-mass 
wind profile from the surface of the star out to the relic H II region: 
\vspace{0.05in}
\begin{equation}
\rho_{\mathrm{W}}(r) = \frac{\dot{m}}{4 \pi r^2 v_{\mathrm{W}}}, \vspace{0.05in}
\end{equation}
where $\dot{m}$ is the mass loss rate due to the wind and $v_{\mathrm{W}}$ is the wind 
speed.  The mass loss rate is \vspace{0.05in}
\begin{equation}
\dot{m} = \frac{M_{\mathrm{tot}}}{t_{\mathrm{MSL}}}, \vspace{0.05in}
\end{equation}
where $M_{\mathrm{tot}}$ and $t_{\mathrm{MSL}}$ are the total mass loss and lifetime 
of the star, respectively. We take $M_{\mathrm{tot}} =$ 0.1 \Ms, $v_{\mathrm{w}} =$ 1000 
km s$^{-1}$, and $T_{\mathrm{W}} =$ 0.01 eV in all our models.  The wind is H and He 
only, with mass fractions of 76\% and 24\%, respectively.  The radius at which the density 
of the wind is joined to the H II region, whose density we take to be 0.1 cm$^{-3}$, varies 
with the radius of each star but is typically 10$^{17}$ - 10$^{18}$ cm.  Density and velocity 
profiles for the wind are visible in Fig.~\ref{fig:PISNprof}.   

\subsection{Ionization of the Wind}

To determine if the progenitor ionizes the wind we use the ZEUS-MP code to model the 
propagation of the ionization front from the surface of the star \citep{wn06,wn08b,wn08a}. 
We center a blue 175 \Ms\ z-series star in the wind on a 1D spherical mesh with 200 
zones and inner and outer boundaries at 7.0 $\times$ 10$^{12}$ cm and 1.0 $\times$ 
10$^{15}$ cm, respectively (the surface of the star and the outer regions of the wind). To 
enhance grid resolution in the densest regions of the wind we use logarithmically ratioed 
zones, where \vspace{0.05in}
\begin{equation}
\frac{\Delta r_{i+1}}{\Delta r_i}  = 1.043. \vspace{0.05in}
\end{equation}
We evolve the I-front with multifrequency UV transport, with 40 bins uniformly partitioned 
in energy from 0.255 - 13.6 eV and 80 logarithmically spaced bins from 13.6 - 90 eV. We 
take the spectrum of the star to be blackbody, normalized to ionizing photon rates, 
surface temperatures, and luminosities from \citet{s02}.

The star easily ionizes the wind on timescales of $\sim $ 10$^4$ yr. The gas temperature 
is 40,000 K near the surface of the star, where rapid ionizations and recombinations in the 
large fluxes and densities there lead to greater heating than at 10$^{15}$ cm, where 
temperatures are $\sim$ 25,000 K.  Because the least massive blue star in our study 
easily ionizes the wind (and has the highest surface wind density), we conclude that the 
winds around all z-series progenitors in our simulations are ionized.

Red u-series progenitors have surface temperatures that are too low to emit ionizing UV 
but still ionize their wind envelopes because they are blue for most of their lives.  They 
become too cool to sustain ionizing flux only in their final few hundred kyr. Recombination 
times in the wind therefore determine its ionization state when the star dies, without the 
need for a transport calculation:
\vspace{0.03in}
\begin{equation}
t_{rec} = \frac{1}{n_e \alpha(T)}. \vspace{0.03in}
\end{equation} 
Here, $n_e$ is the electron number density and $\alpha(T)$ is the recombination rate 
coefficient for hydrogen, which we take to be 2.59 $\times$ 10$^{-13}$ $T_4^{-0.75}$ 
s$^{-1}$, where $T_4$ is the temperature in units of 10$^4$ K.  With the ionized gas 
temperatures we found for the compact 175 \Ms\ star and the densities of the modest 
winds in our study, recombination times vary from 5 - 50 kyr in the vicinity of the star, 
ensuring that the wind is neutral when the star explodes.  For simplicity, we take the 
wind to be neutral around all the stars in our study, so our luminosities at shock 
breakout for z-series stars will be lower limits because neutral envelopes allow less 
flux to escape until they are fully ionized by the breakout pulse.  We note that the 
ionized wind itself has significant luminosity due to recombinations \citep{rz10}, but we 
are primarily interested in the transient flux from the explosion, not the steady emission 
from its envelope.
  
\subsection{SPECTRUM}

To calculate a spectrum from a RAGE profile, we map its densities, temperatures, mass
fractions and velocities onto a new grid in the SPECTRUM code.  SPECTRUM performs a 
direct sum of the luminosity of every fluid element in this discretized snapshot to compute 
the total flux that exits the ejecta along the line of sight at every wavelength. This procedure,
described in detail in FET12, accounts for Doppler shifts and time dilation due to relativistic 
expansion of the ejecta.  We also calculate the intensities of emission lines and attenuation 
of flux along the line of sight with monochromatic OPLIB opacities, thereby capturing both 
limb darkening and the absorption lines imprinted on the flux by intervening material in the 
ejecta and wind.

As explained in FET12, gas densities, velocities, mass fractions and radiation temperatures 
from the finest levels of refinement of the RAGE AMR grid are first extracted and ordered by 
radius into separate files, one variable per file.  In these runs, the profiles have 50,000 radial 
zones and constraints on machine memory and time prevent us from using all of them to 
calculate spectra, so only a subset of the points is mapped onto the SPECTRUM grid.  We 
first sample the RAGE radiation energy density profile inward from the outer boundary to find 
the position of the radiation front, which we take to be where $aT^4$ rises above 1.0 erg/cm$
^3$.  We then find the radius of the $\tau = $ 25 surface by integrating the optical depth due 
to Thomson scattering inward from the outer boundary ($\kappa_{Th} =$ 0.288 for H and He 
gas at primordial composition). This yields the greatest depth in the ejecta from which 
photons can escape because $\kappa_{Th}$ is the minimum opacity the photons would 
encounter.  

The extracted densities, velocities, temperatures and species mass fractions from RAGE are 
then interpolated onto a two-dimensional (2D) grid in $r$ and $\theta$ in SPECTRUM whose 
inner and outer boundaries are 0 and 10$^{18}$ cm, respectively.  Eight hundred uniform 
zones in log radius are allocated from the center of the grid to the $\tau =$ 25 surface.  The 
region between the $\tau =$ 25 surface and the edge of the radiation front is partitioned into 
6200 uniform zones in radius. The wind between the front and the outer boundary of the grid 
is then divided into 500 uniform zones in log radius, for a total of 7500 radial bins.  The data 
in each of these new radial bins is mass-averaged to ensure that SPECTRUM captures very 
sharp features from the RAGE profile.  The grid is uniformly discretized into 160 bins in $\mu 
=$ cos$\, \theta$ from -1 to 1.  Our choice of mesh yielded good convergence in spectrum 
tests, fully resolving regions of the flow from which photons can escape the ejecta and only 
lightly sampling those from which they cannot.

Summing the luminosities at each wavelength in one spectrum yields the bolometric luminosity 
of the SN at that moment. Many such luminosities computed over a range of times constitutes 
the light curve of the explosion.  We sample the light curve with 200 - 340 spectra that are 
logarithmically distributed in time out to 3 yr.  

\section{PI SN Blast Profiles, Light Curves and Spectra}

\begin{figure}
\plotone{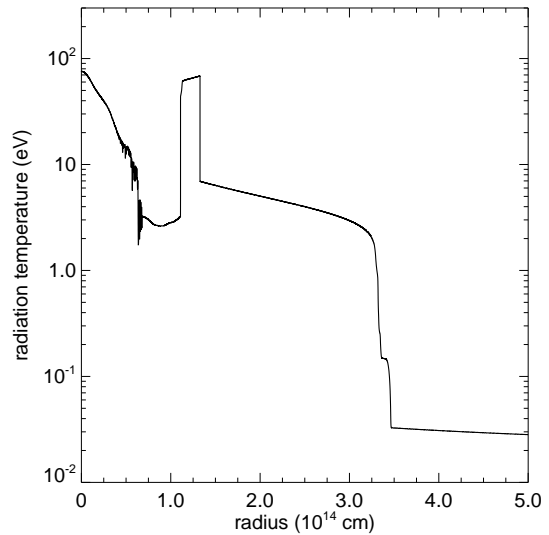}
\caption{The highly radiative shock in the u225 model at 1.59 $\times$ 10$^{5}$ s.  The shock is 
at 1.2 $\times$ 10$^{14}$ cm, deep inside the star.  Photons from the shock cannot reach the 
surface of the star from this depth because of Thomson scattering in its upper layers.  The 0.2 eV 
plateau in the temperature at the stellar surface at 3.2 $\times$ 10$^{14}$ cm is optical radiation 
from the star propagating out into the wind before shock breakout.
\vspace{0.1in}}
\label{fig:radshk}
\end{figure}

\begin{figure*}
\plottwo{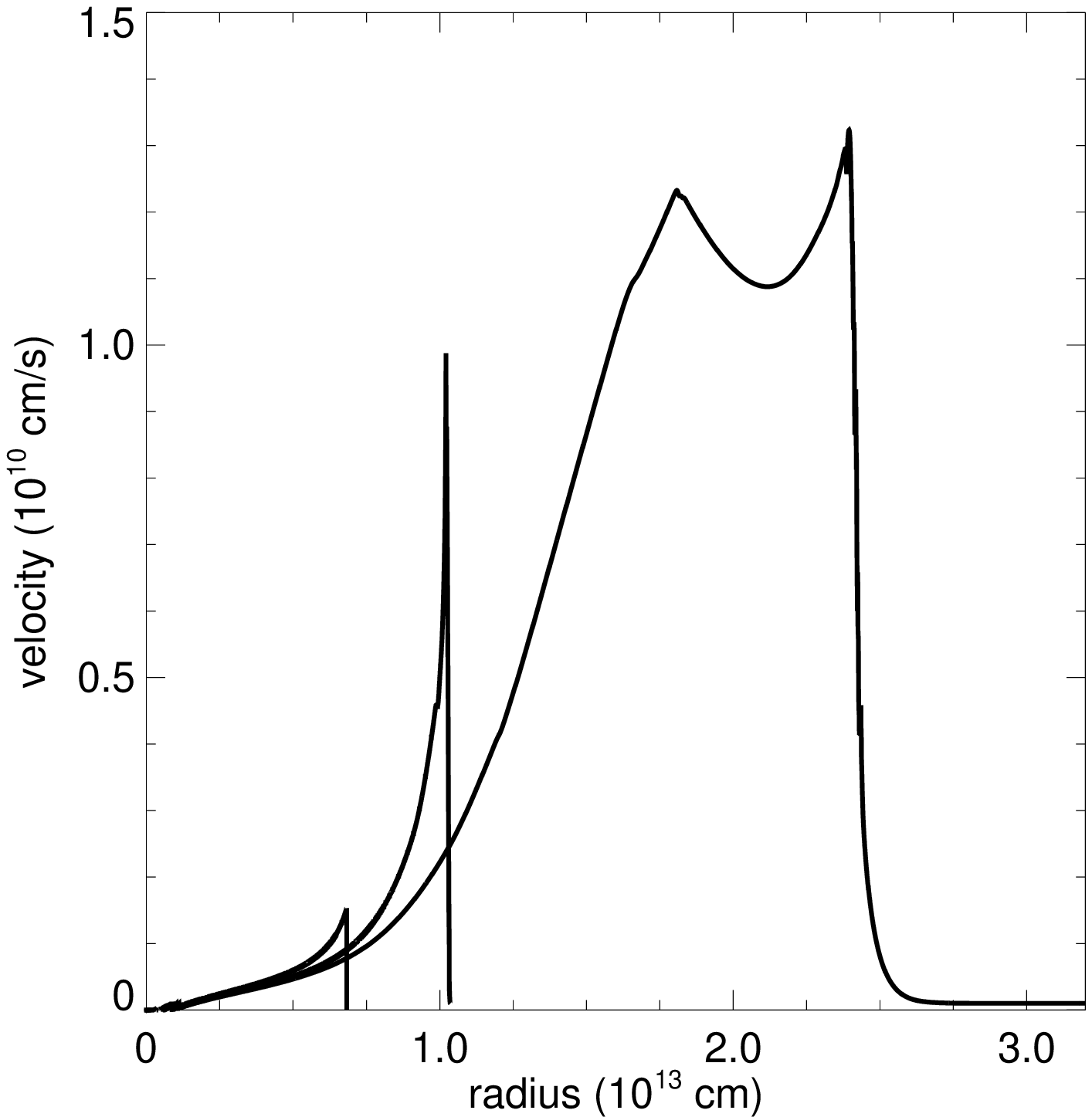}{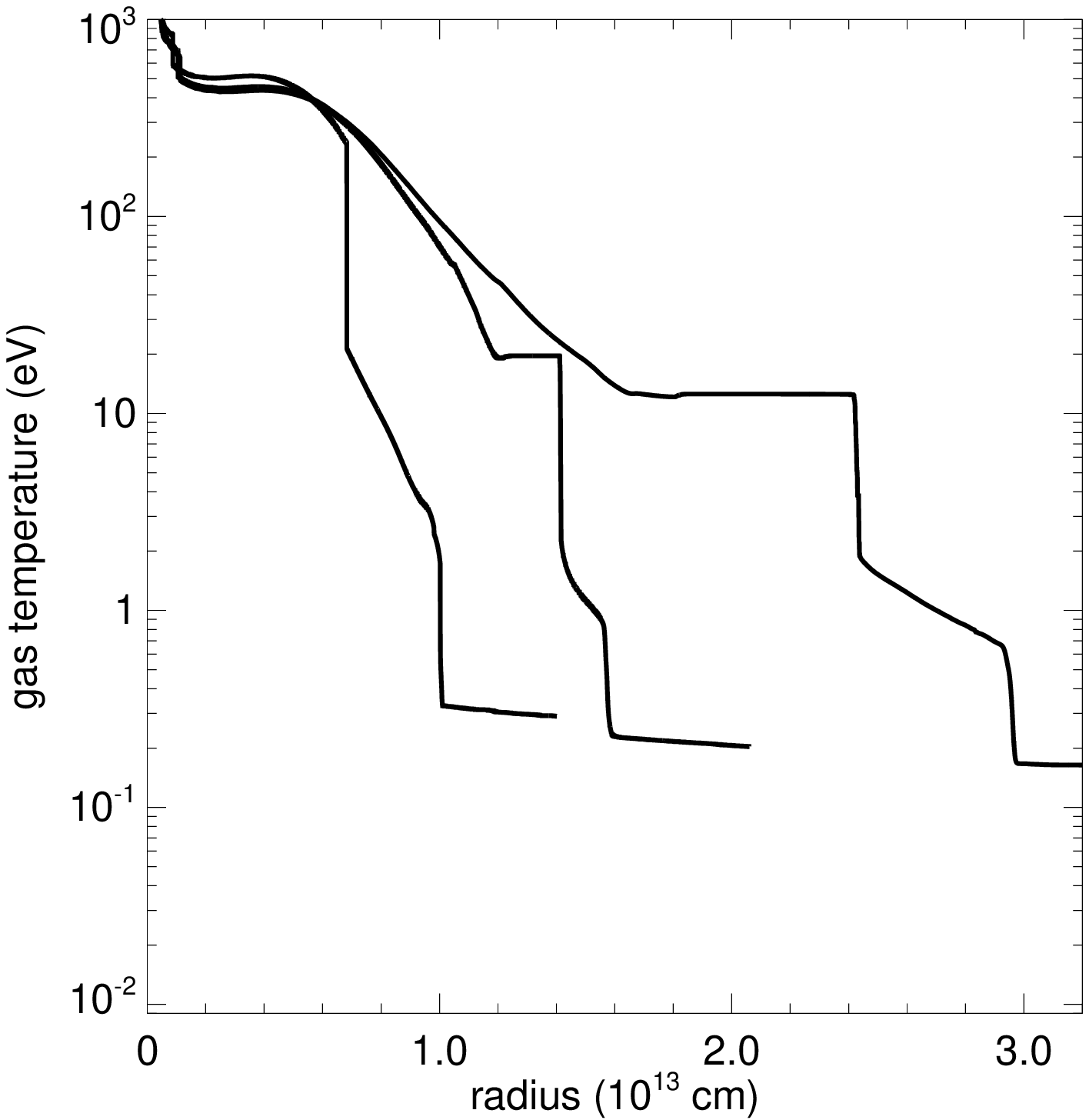}
\caption{Shock breakout in the z225 explosion.  Left panel:  velocities, from left to right, at 8309 s, 9325 
s, and 9962 s.  Shock breakout through the surface of the star is evident in the abrupt jump in velocity in 
the second profile, and the continued acceleration of the shock down the $r^{-2}$ density profile of the 
wind is visible in the third profile.  Right panel:  gas temperatures at 8309 s, 9502 s, and 9962 s from left 
to right. The radiation breakout pulse is the flat 20 eV plateau in gas temperature at 1.2 $\times$ 10$^{
13}$ cm at 9502 s. As the shock expands and cools the spectrum of the breakout pulse softens, which is 
why the gas temperature in its wake has fallen to $\sim$ 10 eV at 9962 s.    
\vspace{0.1in}}
\label{fig:breakout}
\end{figure*}

\begin{figure*}
\plottwo{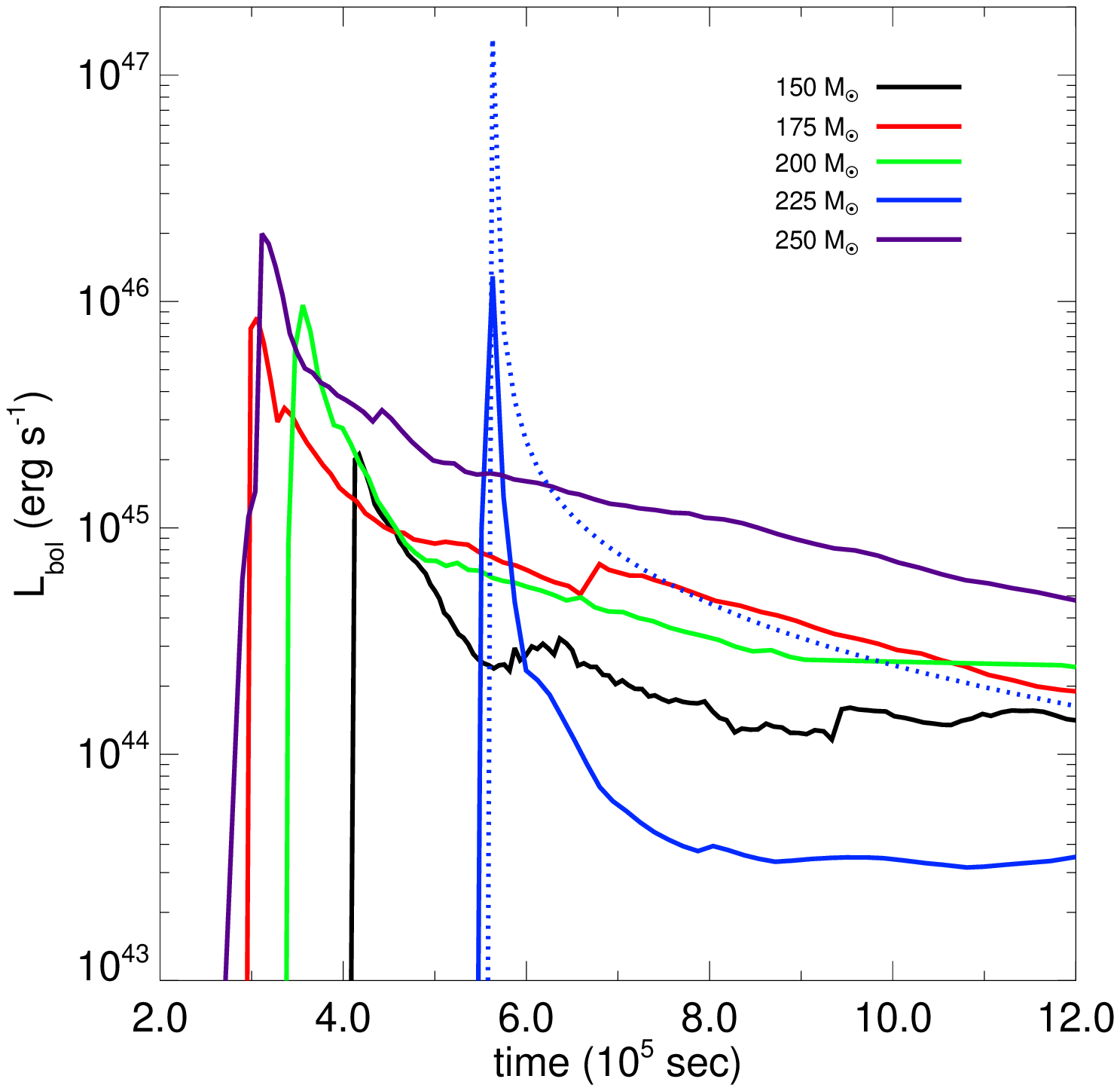}{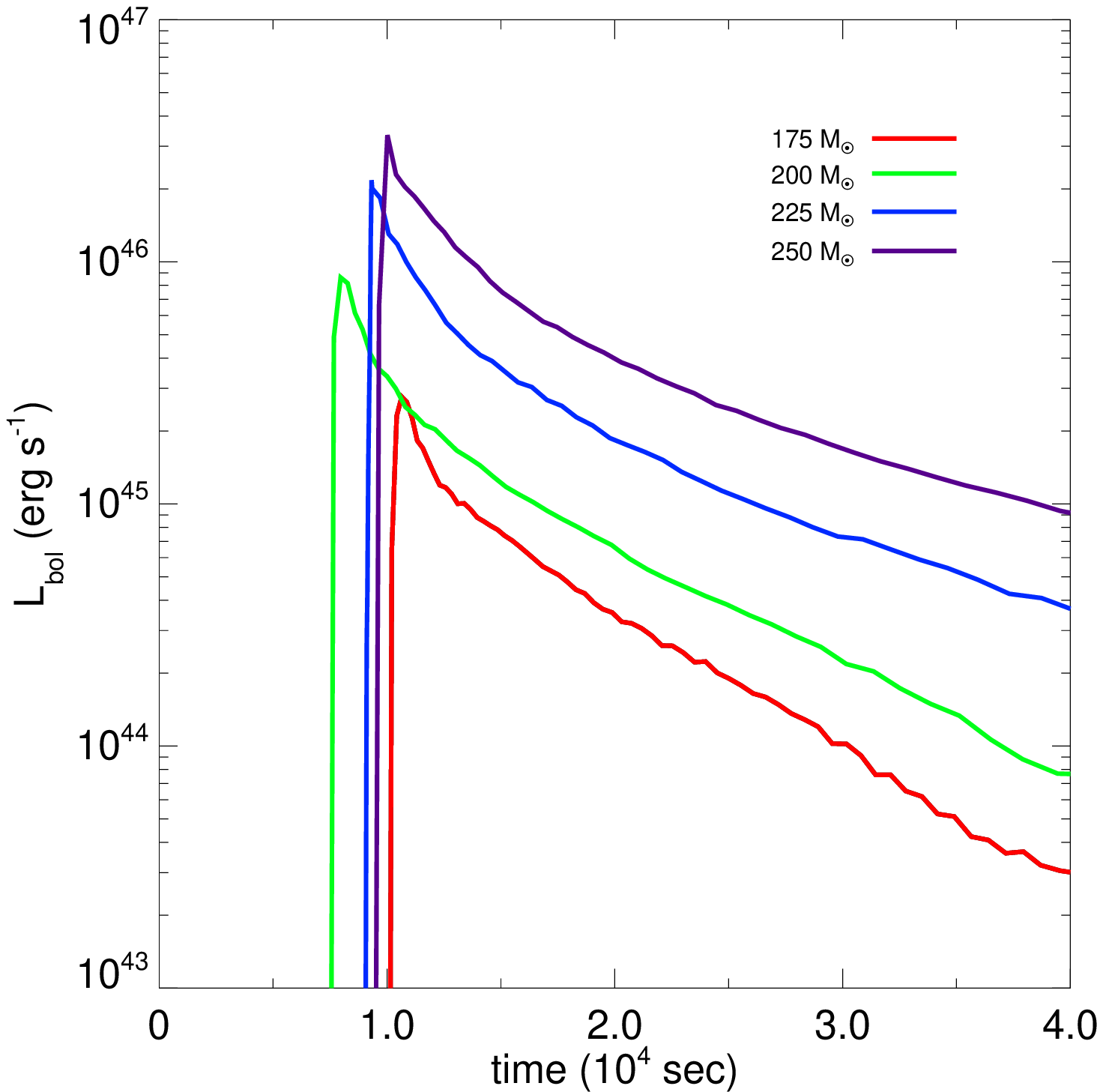}
\caption{Bolometric luminosities.  Left panel:  u-series PI SNe.  The dotted blue line is the blackbody 
approximation to the u225 light curve (equation 11).  Right panel:  z-series PI SNe.    
\vspace{0.1in}}
\label{fig:breakoutLC}
\end{figure*}

\begin{figure*}
\plottwo{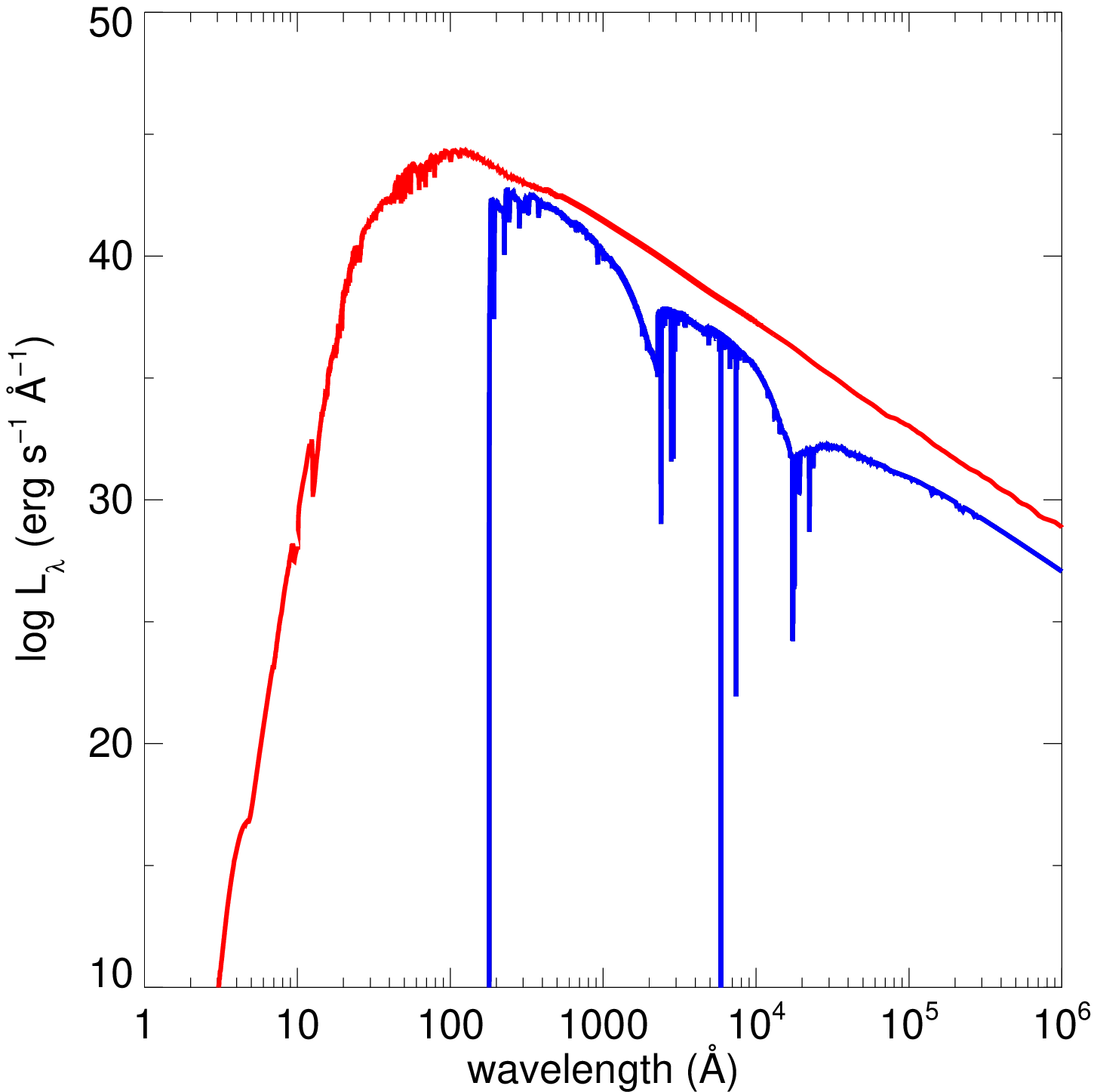}{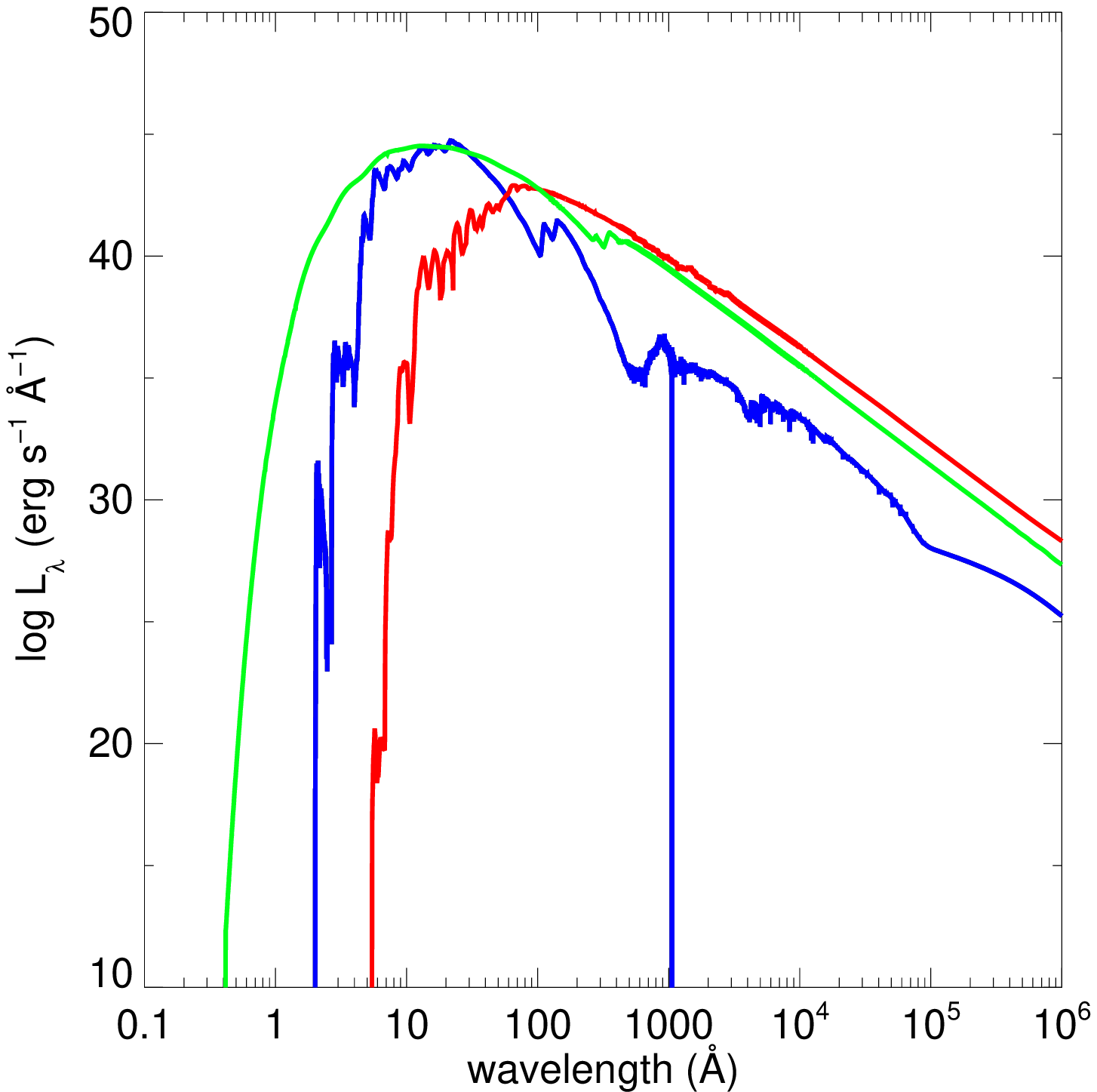}
\caption{Spectra of the breakout transient.  Left panel: u250 explosion.  Blue: 2.90 $\times$ 10$^5$ s; 
red:  3.11 $\times$ 10$^5$ s.  Right panel: z250 explosion.  Blue:  9652 s;  green:  1.40 $\times$ 10$
^4$ s;  red:  4.42 $\times$ 10$^4$ s.
\vspace{0.1in}}
\label{fig:breakoutSP}
\end{figure*}

The onset of explosive burning drives a strongly radiating shock from the O layer into the upper layers 
of the star, as we show in Fig.~\ref{fig:radshk}.  At this point the shock is not visible to an external 
observer because its photons are trapped by $e^-$ scattering in the intervening layers, so they are 
simply advected outward by the fluid flow.  The SN becomes visible when the shock breaks through
the surface of the star.  Shock breakout has been the subject of numerous analytical studies \citep{
cg74,mm99,ns10,Piro10,Katz12} and numerical studies \citep{Ens92,Blinn00,Tomin09,Tolstov10,
kasen11,bt11,tbn13} in the past thirty years but has not been observed until more recently \citep[e.g.,
][]{sod08,gez08,sch08,dwek08}.

\subsection{Shock Breakout}

When the shock reaches the surface of the star it abruptly accelerates to large velocities in the steep 
density gradient there, as we show in the left panel of Fig.~\ref{fig:breakout}.  The acceleration 
heats the shock, and it releases an intense burst of photons upon being exposed to the low density 
IGM, as shown in the right panel of Fig.~\ref{fig:breakout}.  The advancing radiation front is visible 
as the 20 eV plateau in gas temperature at 1.2 $\times$ 10$^{13}$ cm at 3990 s that is 10 eV at 2.0 
$\times$ 10$^{13}$ cm at 4550 s.  The plateau temperatures are those to which the radiation front 
heats the wind as it passes through it; the shock that is emitting the radiation is much hotter, as we 
show below.  The temperature of the plateau falls as the shock expands, cools and its spectrum 
softens.  Note that there are serious departures from self-similarity in the velocity profiles at late 
stages of breakout (third plot in the left panel of Fig.~\ref{fig:breakout}) because of significant 
coupling between radiation and gas in the shock.  This is one reason Sedov-Taylor profiles are not 
good solutions for shock breakout \citep{fwf10} and why radiation transport is required to model the 
flow.

We show bolometric luminosities for the breakout transient for the u-series and z-series PI SNe in 
the left and right panels of Fig.~\ref{fig:breakoutLC}, respectively.  For comparison, we show the 
blackbody approximation to the luminosity of the shock for the u225 explosion, \vspace{0.1in}
\begin{equation}
L = 4 \pi r^2 \sigma T^4, \label{eq:BB} \vspace{0.1in}
\end{equation} 
where $\sigma$ is the Stefan-Boltzmann constant and the temperature $T$ of the shock is taken at the 
$\tau_{Th} = $ 1 surface, where $\kappa_{Th} = $ 0.288.  If photons of all energies broke through the 
surface of the star at the same time, the duration of the initial transient would be comparable to the light
crossing time of the star, since photons emitted from its poles and its equator would reach an observer 
at times that differ by the time it takes light to cross the star.  Given that u-series stars have radii of a few 
10$^{14}$ cm, their breakout transients would last 1 - 2 hr, which is consistent with the width of the $4 \pi 
r^2 \sigma T^4$ approximation to the luminosity of the pulse. In reality, the transient is smeared out over 
longer times because radiation remains strongly coupled to the shock past breakout (as shown in the third 
velocity profile in Fig.~\ref{fig:breakout}).  The resulting pulse is dimmer but longer.  The width of the 
transient is much greater for giant red u-series stars because of their larger radii, lower surface densities, 
and thus broader regions from which photons break free of the shock. 

\begin{figure*}
\plottwo{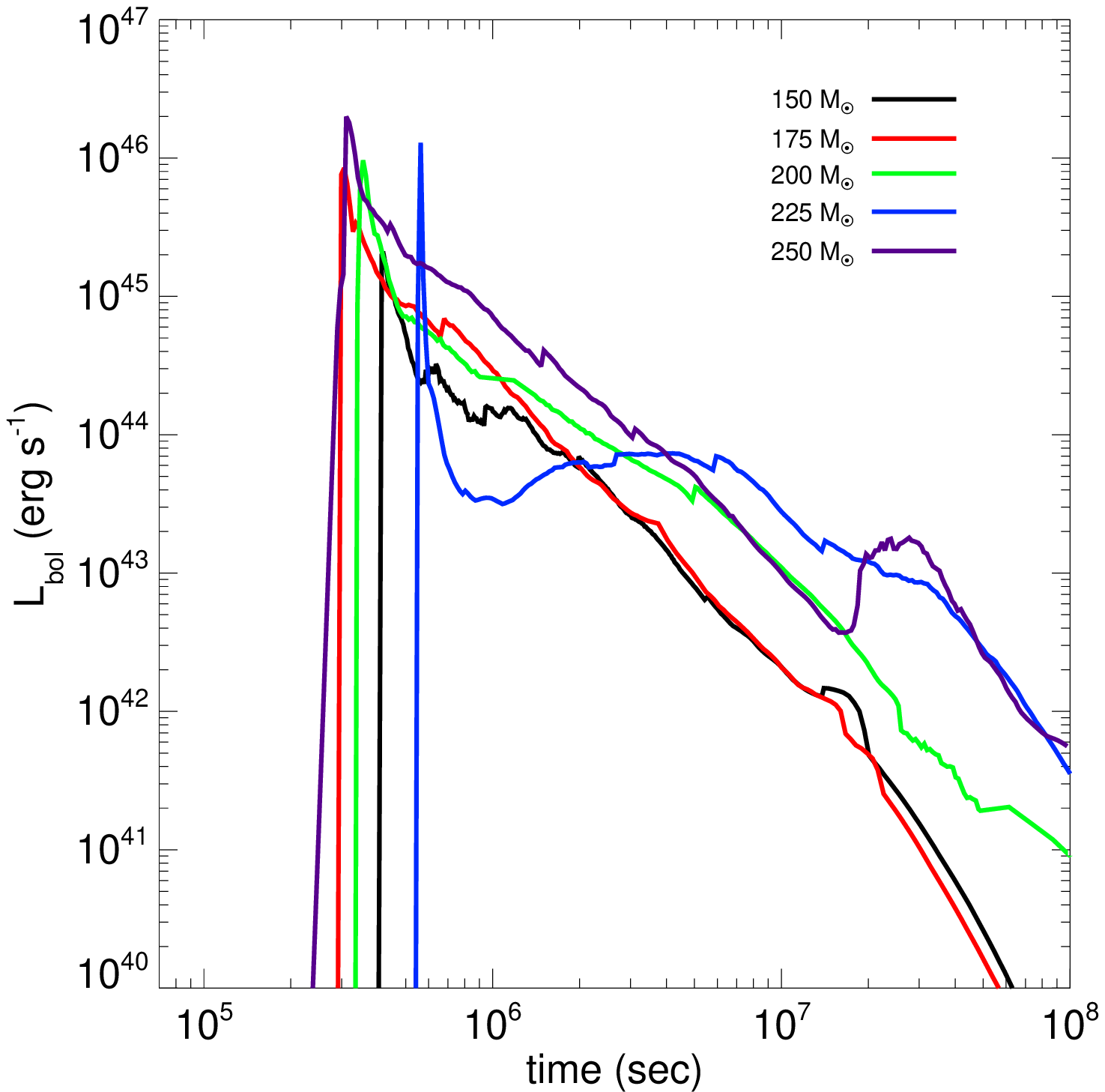}{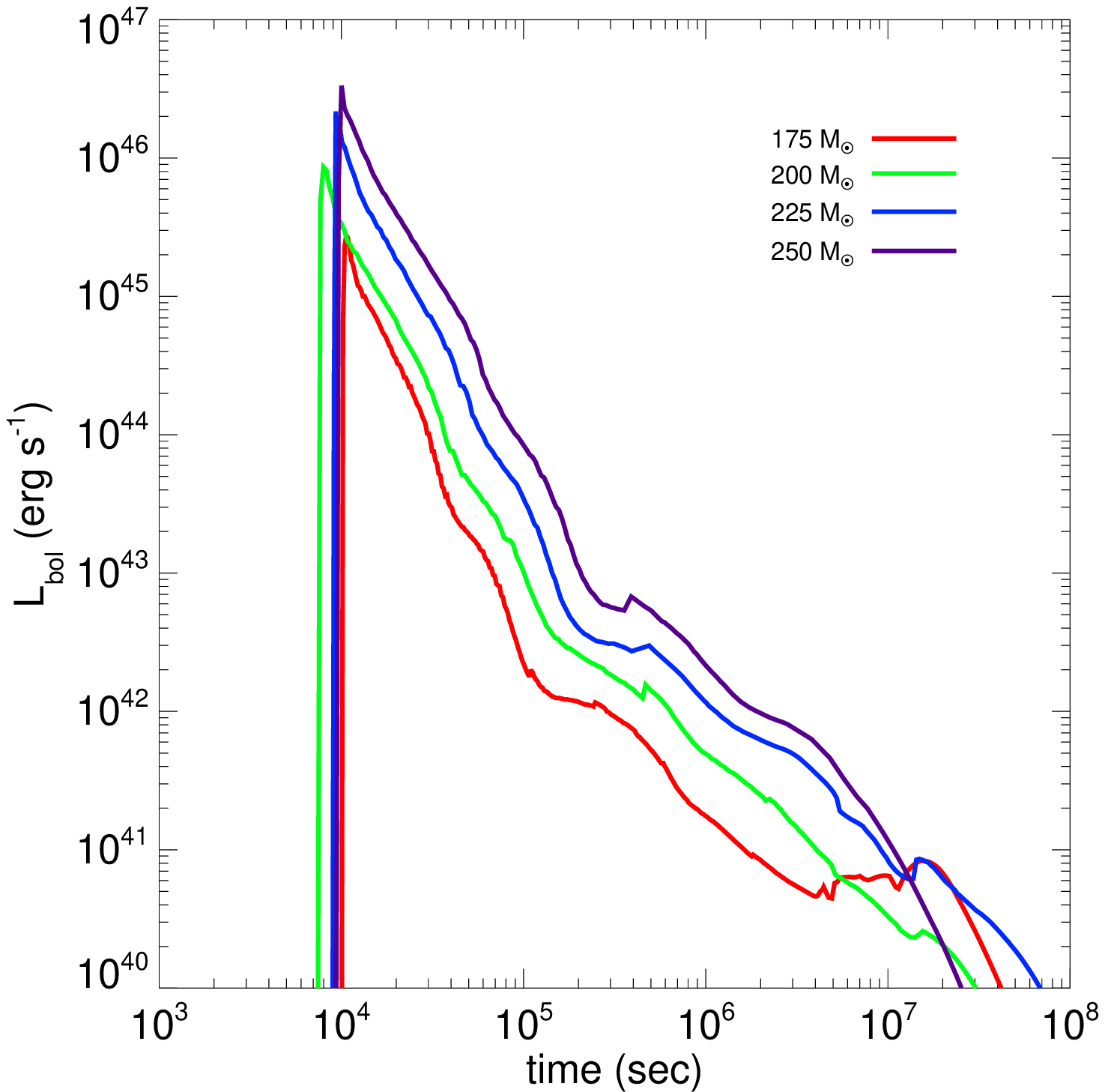}
\caption{Source frame bolometric luminosities for all 9 PI SNe out to 3 yr.  Left panel:  u-series.  Right 
panel:  z-series. The general trend of higher luminosity with progenitor mass in each series is evident, 
as is the fact that u-series explosions are brighter than z-series SNe of equal progenitor mass.  The 
resurgence in luminosity at $\sim$ 10$^7$ s in most of the SNe coincides with the descent of the 
photosphere of the shock into the hot \Ni\ layer of the ejecta. 
\vspace{0.1in}}
\label{fig:LC}
\end{figure*}

\begin{figure}
\plotone{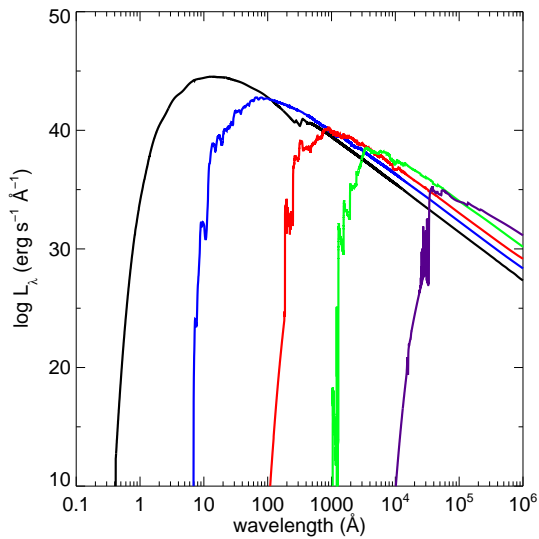}
\caption{Spectral evolution of the z250 PI SN.  Fireball spectra at 1.40 $\times$ 10$^4$ s (black), 4.77 
$\times$ 10$^4$ s (blue),  2.10 $\times$ 10$^5$ s (red),  1.95 $\times$ 10$^6$ s (green), and 3.39 $
\times$ 10$^7$ s (purple).\vspace{0.1in}}
\label{fig:z250sp}
\end{figure}

We show spectra for the breakout pulse for u250 and z250 PI SNe in the left and right panels of Figure 
\ref{fig:breakoutSP}, respectively.  The z250 transient is mostly x-rays and the u250 pulse is both x-rays 
and hard UV.  The spectrum of the u250 transient is softer because the progenitor is much larger, and 
the shock has done more $PdV$ work on its surroundings and cooled more by the time it breaks out of 
the star.  The ionization of the surrounding wind envelope is particularly evident in the z250 breakout 
spectra, as the prominent absorption features at 9652 s are mostly gone by 1.40 $\times$ 10$^4$ s. By 
the end of shock breakout the surrounding wind is completely ionized by the radiation pulse and the 
spectrum of the shock has essentially a blackbody profile.  

\begin{figure*}
\plottwo{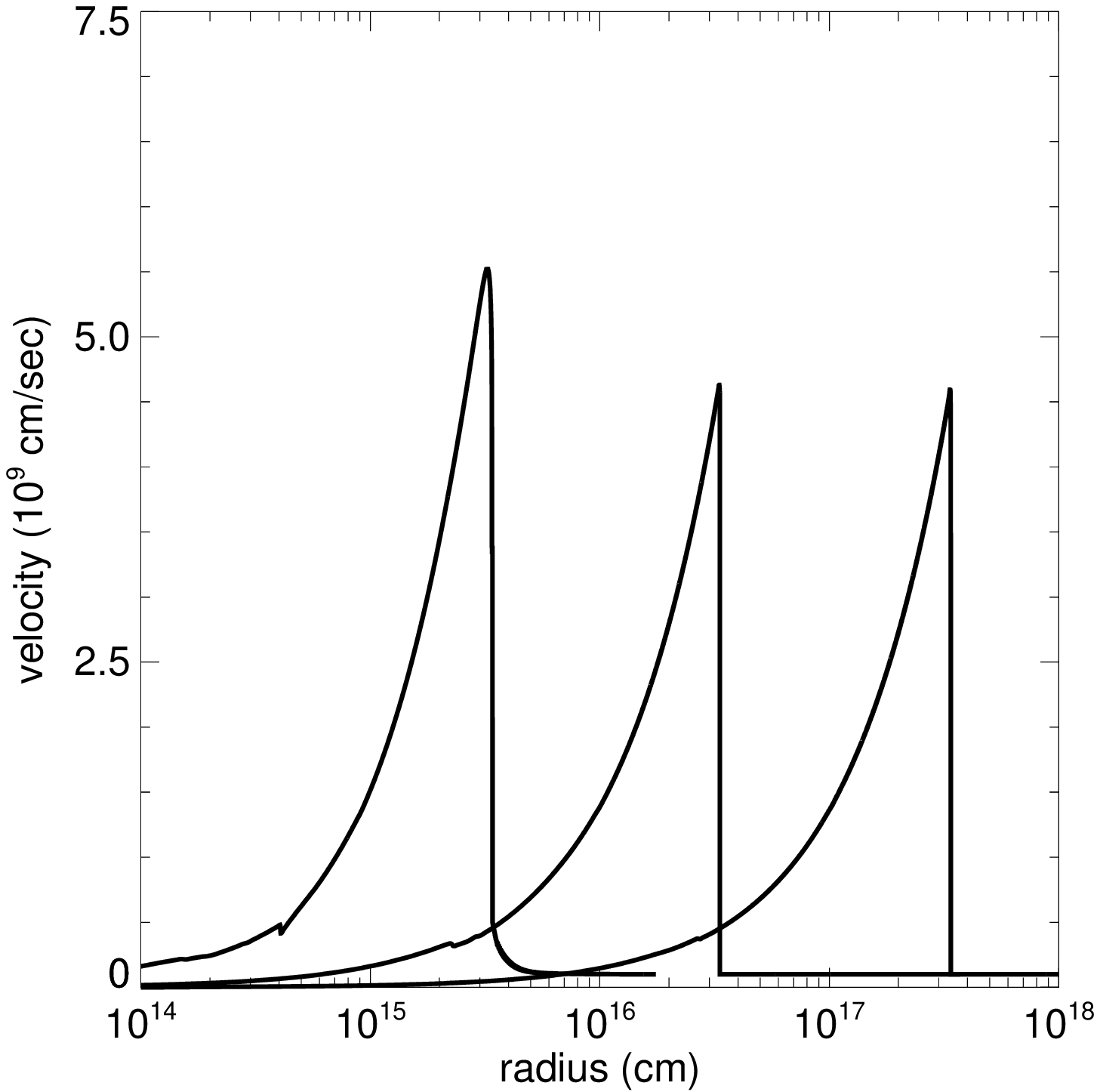}{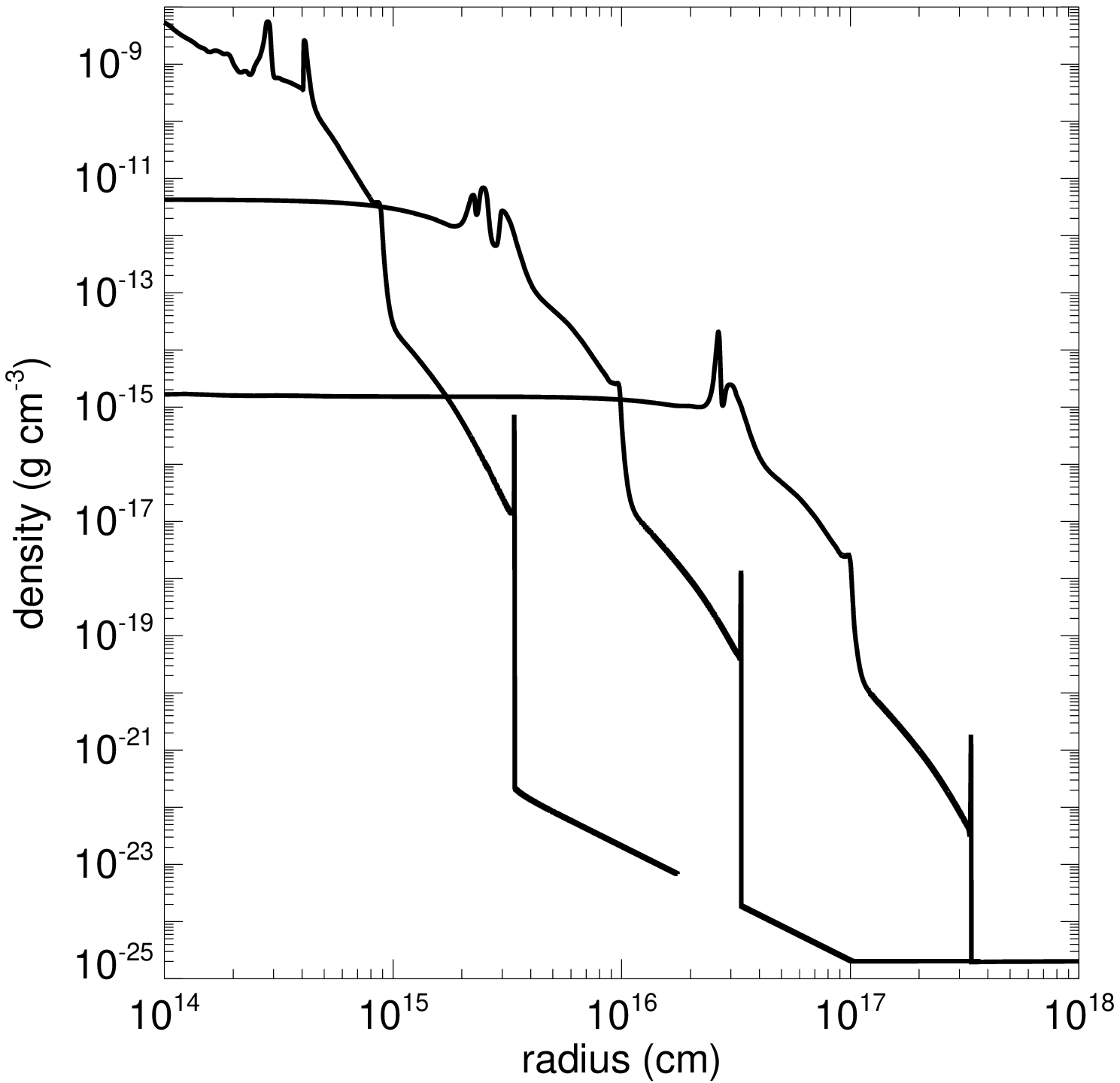}
\caption{Intermediate to late-time hydrodynamical evolution of the u250 PI SN.  Left panel:  velocities at 
8.36 $\times$10$^5$ s, 7.41 $\times$ 10$^6$ s,  and 7.36 $\times$ 10$^7$ s (left to right).  Right panel: 
density profiles at the same times from left to right.
\vspace{0.1in}}
\label{fig:hydro}
\end{figure*}

Peak bolometric luminosities vary from 8 $\times$ 10$^{45}$ to 3 $\times$ 10$^{46}$ erg/s, or $\sim$ 200 
times the luminosity of our galaxy.  These peak luminosities are consistent with those of \citet{kasen11},
which were computed with the Kepler and SEDONA codes utilizing the Lawrence Livermore OPAL opacities 
\citep{opal1,opal2}\footnote{http://rdc.llnl.gov}. There is a general trend of greater luminosity with explosion 
energy and later breakout times with progenitor mass, since stellar radii increase with mass.  Although the 
shock has a much smaller radius at breakout in the z-series than in the u-series, it has about the same total 
luminosity because it is hotter.  This can be seen from how the luminosity $L$ of a blackbody scales with 
radius and temperature in equation \ref{eq:BB}.  Typical shock temperatures at the $\tau = 1$ surface at 
breakout are $\sim$ 500 eV in the z-series stars and $\sim$ 50 eV in the u-series stars, which roughly 
compensates for the factor of 10 in radius between the two progenitors.  

However, we note that the luminosity of the shock is not well approximated by that of a blackbody and that it 
is roughly an order of magnitude below that predicted by equation \ref{eq:BB}, as we show in 
Fig.~\ref{fig:breakoutLC}\footnote{Because of $e^-$ scattering in the photosphere of the ejecta, the luminosity
of the shock is better modeled by $L = 4 \pi \epsilon r^2 \sigma T^4$, where $\epsilon < 1$ is the correction to 
the blackbody luminosity due to scattering.}.  Treating SN shocks as blackbodies in general can lead to 
overestimates in luminosity of an order of magnitude or more at breakout, as discussed in \citet{fwf10}.  One 
aspect of shock breakout that is unique to our study is that it is redshifted from 1 - 2 hr in duration at $z \sim$ 
20 to a day or more in the observer frame.  This would enhance the probability of detecting ancient SNe if not 
for the fact that the transient is completely absorbed by the neutral IGM at high redshift.

\subsection{Intermediate / Late Stages of the SN}

We show bolometric light curves out to three years for the u-series and z-series SNe in the left and right 
panels of Fig.~\ref{fig:LC}, respectively.  PI SN luminosities are powered at early times by the conversion 
of kinetic and radiation energy into thermal energy by the shock, so they are far brighter than Type Ia and II 
SNe at this stage because they have much higher explosion energies. At later times their luminosity comes 
mostly from radioactive decay, and they are much brighter than other SNe because they synthesize more 
\Ni:  up to 40 \Ms\ compared to 0.4 -- 0.8 \Ms\ and $<$ 0.3 \Ms\ in Type Ia and Type II SNe, respectively.  
They are brighter for longer times (3 years instead of 3 - 6 months for Type Ia and II SNe) because 
radiation diffusion timescales in their more massive ejecta are much longer:
\begin{equation}
t_d \sim \kappa^{\frac{1}{2}} {M_{ej}}^{\frac{3}{4}} E^{-\frac{3}{4}}. \vspace{0.1in}
\end{equation}
Here, $\kappa$ is the average opacity of the ejecta, $M_{ej}$ is the mass of the ejecta, and $E$ is the 
explosion energy.  The luminosity generally rises with progenitor mass within each series because the 
explosion energy and \Ni\ mass increase with stellar mass in both red and blue stars.  All five u-series 
explosions exhibit a slower, more protracted decay in luminosity out to 4 months than the z-series SNe 
and are more than an order of magnitude brighter over this interval.  These profiles are consistent with 
those of Type II plateau SNe, whose progenitors are also thought to be red giants with extended 
envelopes.  Their greater luminosities over this interval are due to the fact that the u-series SN shocks 
are twice as hot as the z-series shocks, in part because u-series explosions create more \Ni\ than 
z-series SNe for progenitors of equal mass (see Table \ref{tab:table1}).  

There is prominent bump in luminosity at 2 - 4 months that lasts for about a year in the u225 and 
u250 light curves.  This rebrightening happens when photons diffusing out from the \Ni\ layer deep 
in the ejecta begin to reach the photosphere and escape into the IGM. The magnitude and duration 
of the bump is proportional to the \Ni\ mass, which explains its absence in the u150 and u175 SNe,
which create less than a tenth of the \Ni\ formed in the u225 SN.  The escape of photons from the 
\Ni\ layer is visible in the u200 light curve as the inflection upward in luminosity at about 1 month.  
Such inflections are also present in all four z-series explosions, which do rebrighten somewhat but 
much less so than the u-series.  The z-series SNe are dimmer at these intermediate times because 
they create less \Ni\ than u-series explosions and because their \Ni\ remains deeper in the ejecta.

We show the evolution of the spectra for the z250 PI SN from shock breakout to 1 yr in 
Fig.~\ref{fig:z250sp}.  Two physical processes govern the evolution of the spectrum over time.  
First, as the fireball expands it cools, and its spectral cutoff advances to longer wavelengths over time.  
Second, the wind envelope that was ionized by the breakout pulse begins to recombine and absorb 
photons at the high energy end of the spectrum, as evidenced by the flux that is blanketed by lines at 
the short-wavelength limit of the spectrum.  At later times flux at longer wavelengths slowly rises due 
to the expansion of the surface area of the photosphere.  Over the wavelength scale of this plot the 
many thousands of lines captured by the LANL OPLIB opacities in our SPECTRUM calculation are 
blended together in the prominent jagged spectral features throughout the spectrum.  Given how line 
blanketing by both the ejecta and the wind shear off the spectrum at short wavelengths, it is clear that 
the common practice of fitting blackbodies to light curves to approximate spectra overestimates flux at
high energies from which many photons are eventually redshifted into the NIR in the observer frame.

\begin{figure*}
\begin{center}
\begin{tabular}{cc}
\epsfig{file=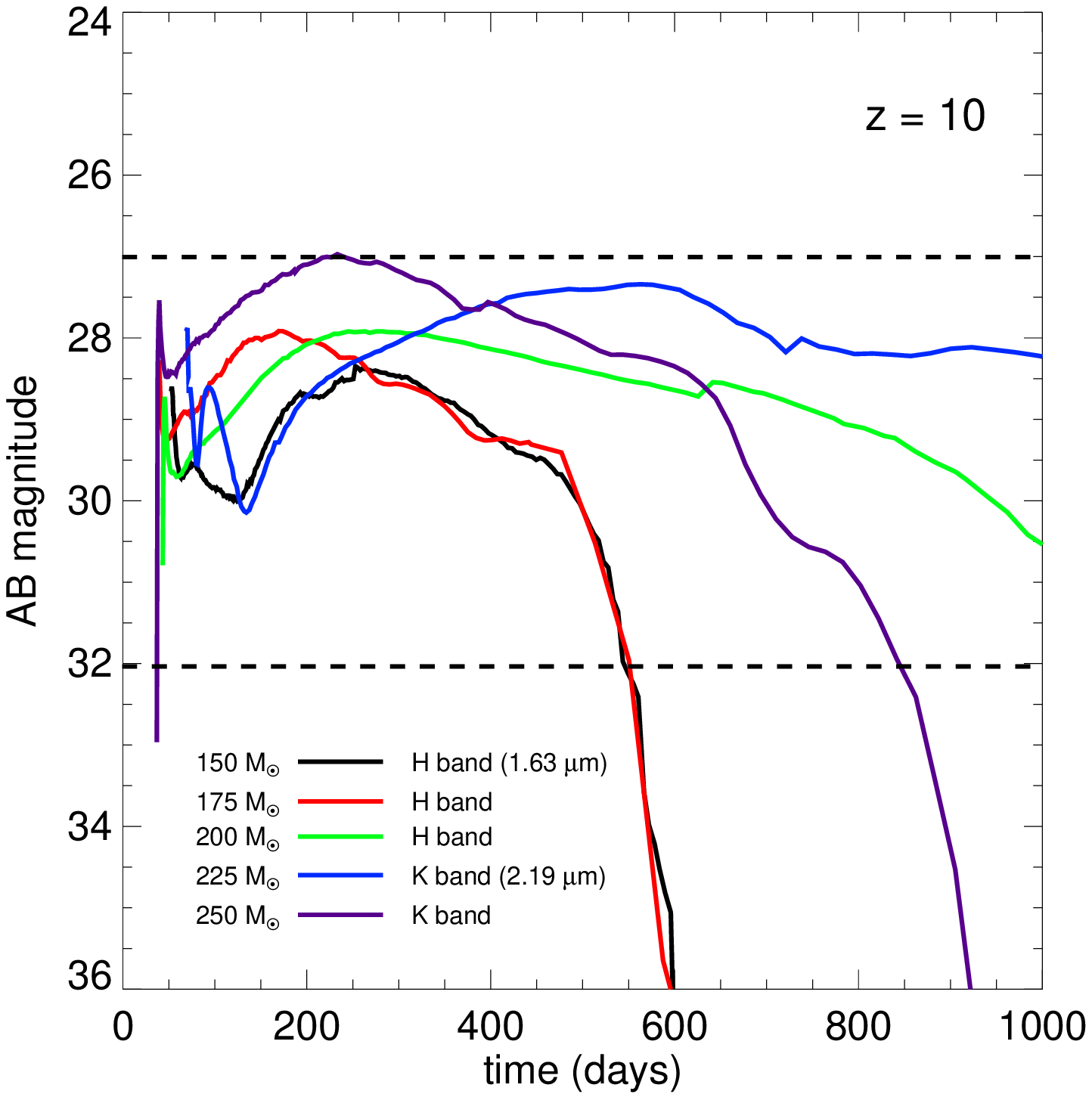,width=0.45\linewidth,clip=} & 
\epsfig{file=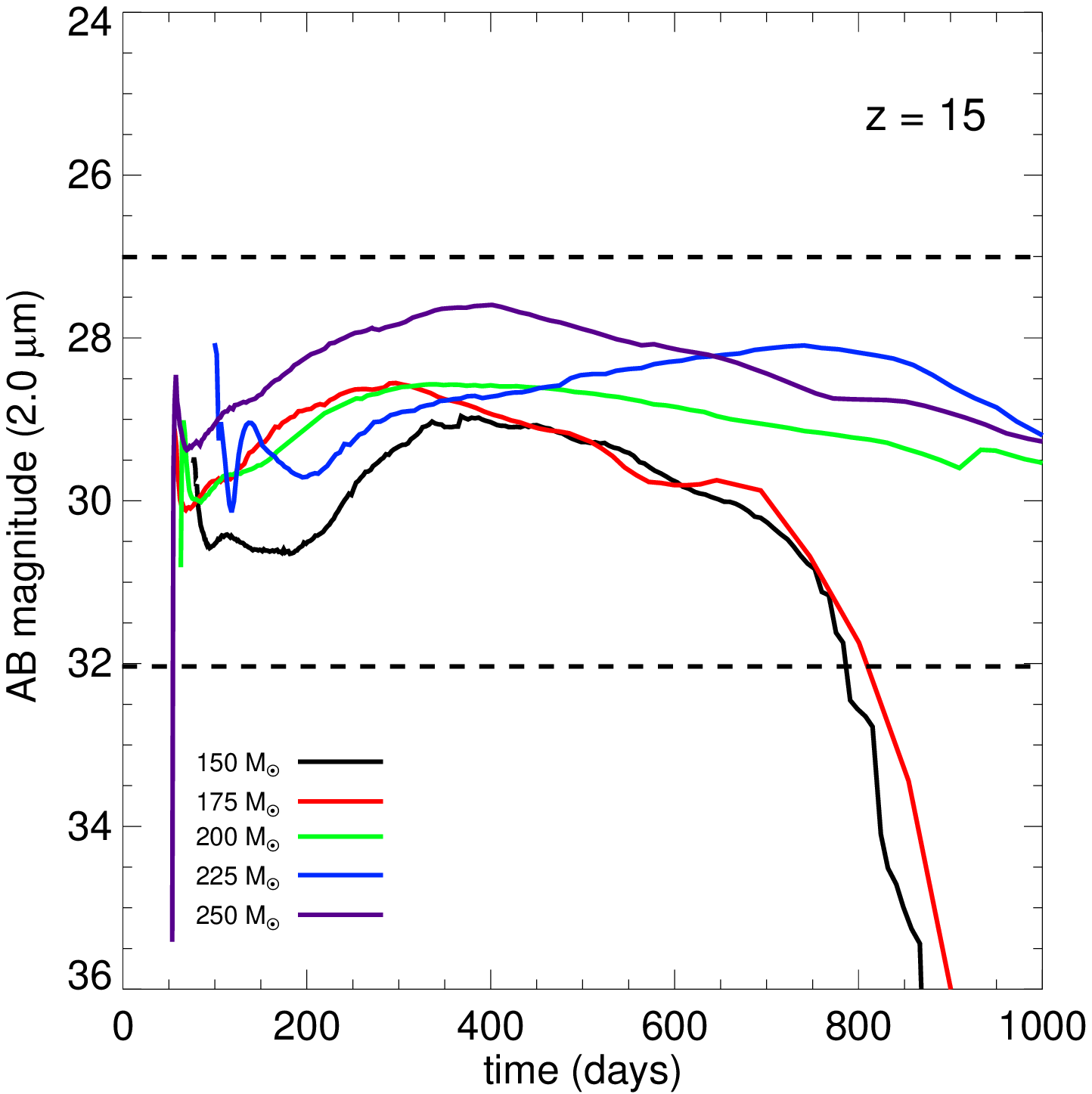,width=0.45\linewidth,clip=} \\
\epsfig{file=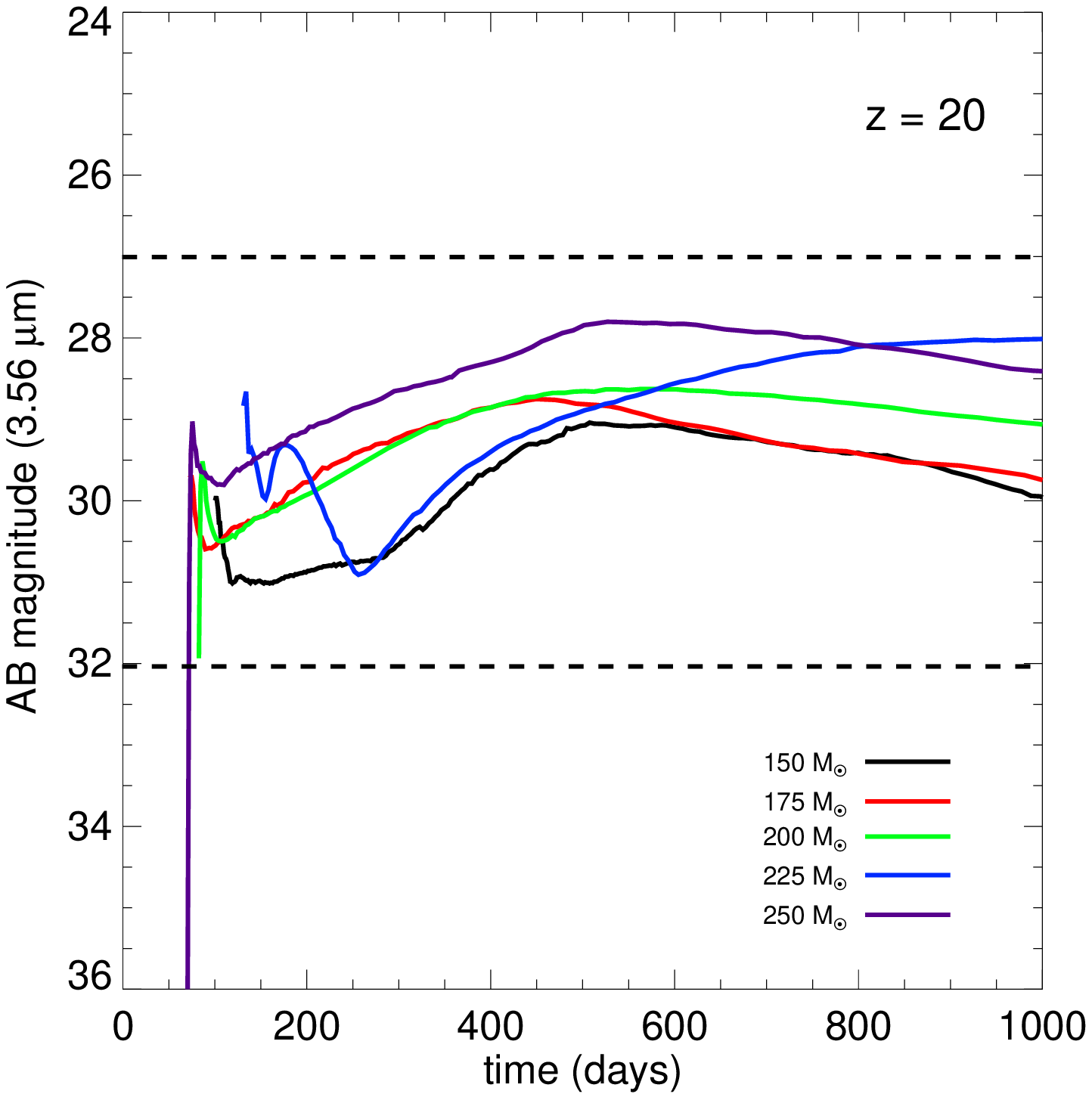,width=0.45\linewidth,clip=} &
\epsfig{file=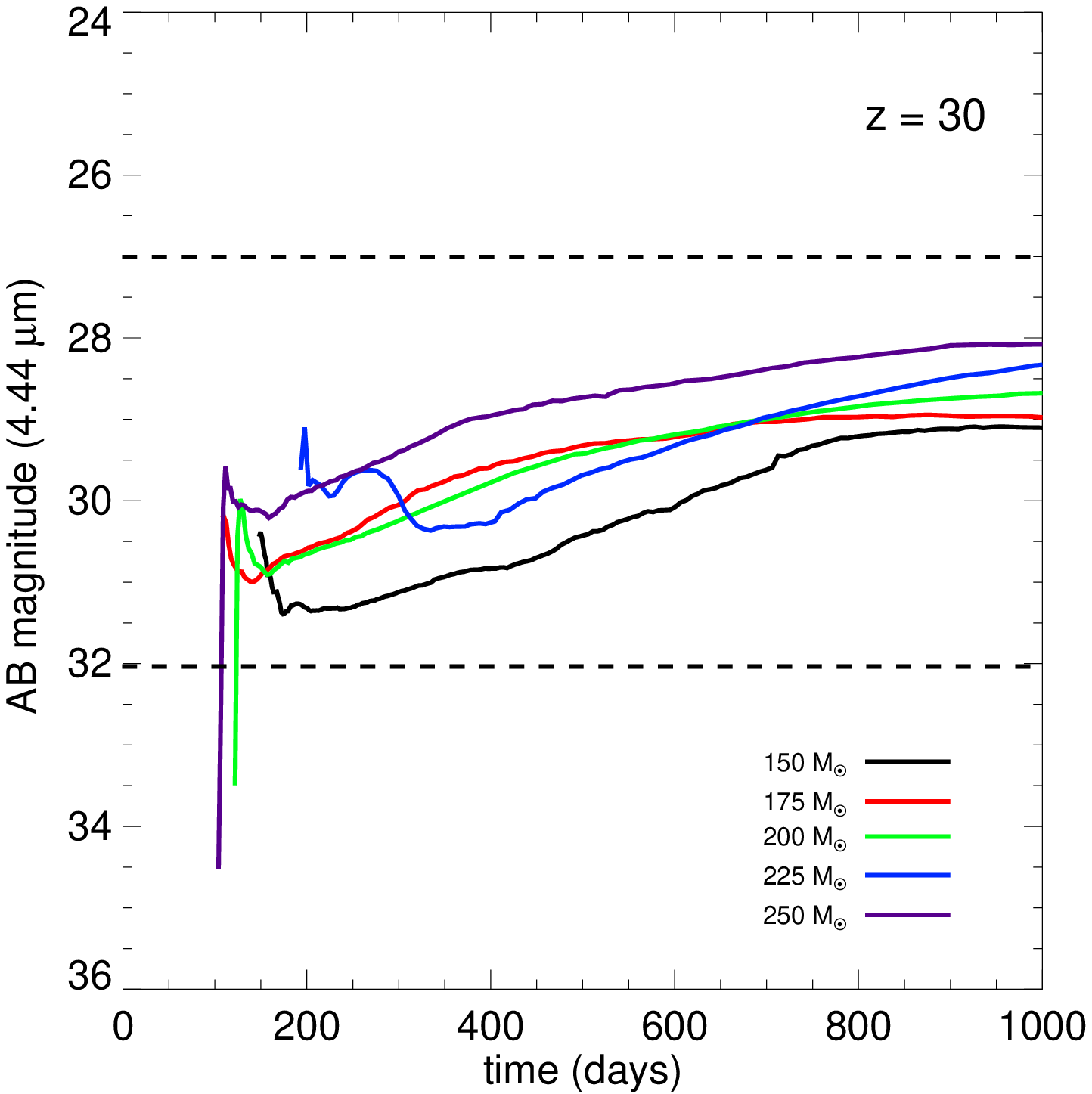,width=0.45\linewidth,clip=}
\end{tabular}
\end{center}
\caption{\textit{JWST} NIRCam light curves for the 5 u-series PI SNe. Redshifts ($z =$ 10, 15, 20, 30) 
are noted in the upper right corner of each panel.  The wavelength of the optimum \textit{JWST} filter 
at each redshift is noted on the y-axis labels and the times on the x-axes are in the observer frame. 
The dashed horizontal lines at AB mag 32 and 27 are \textit{JWST} and WFIRST detection limits, 
respectively.}
\label{fig:uLC}
\end{figure*}

\begin{figure*}
\begin{center}
\begin{tabular}{cc}
\epsfig{file=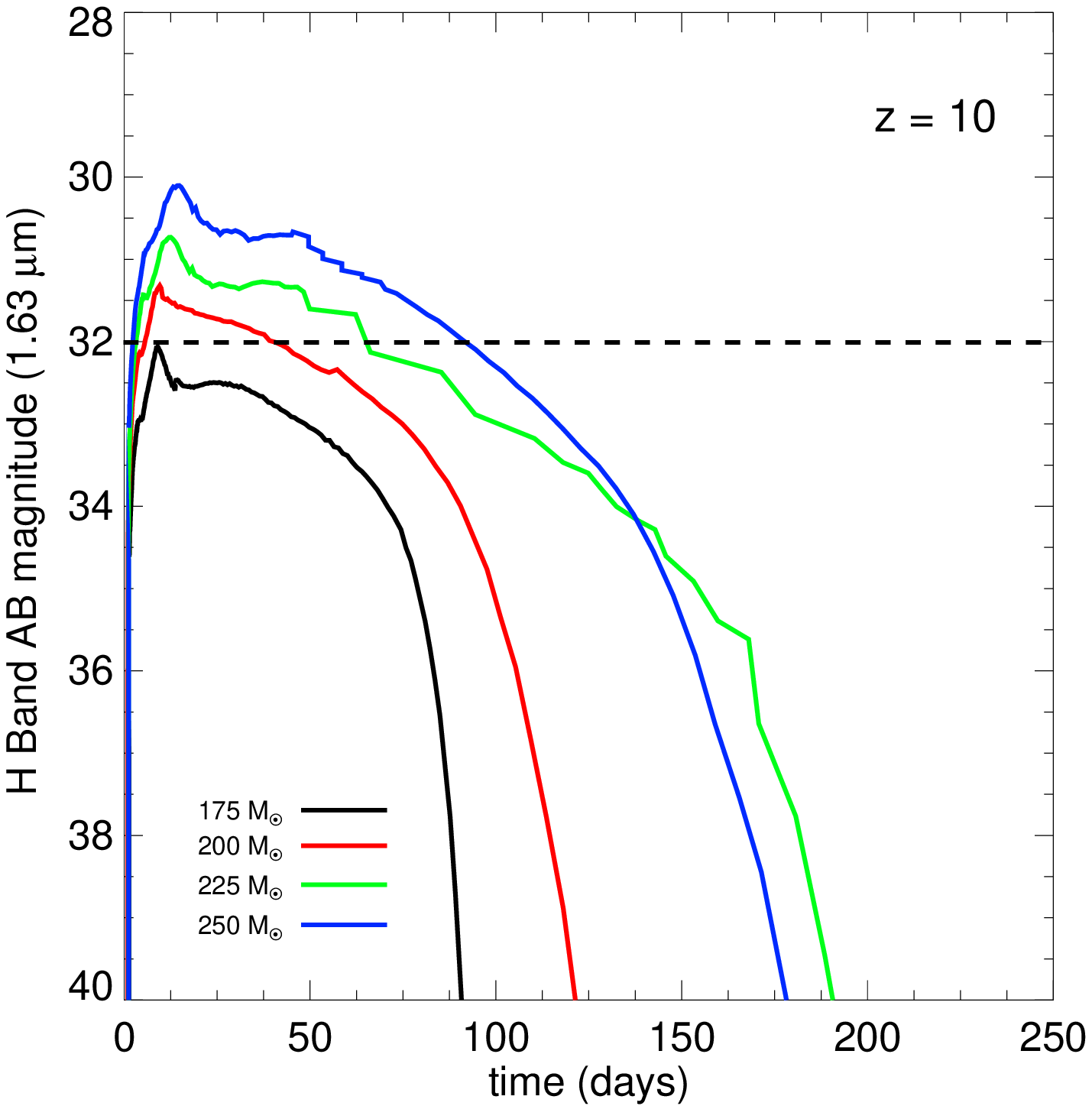,width=0.45\linewidth,clip=} & 
\epsfig{file=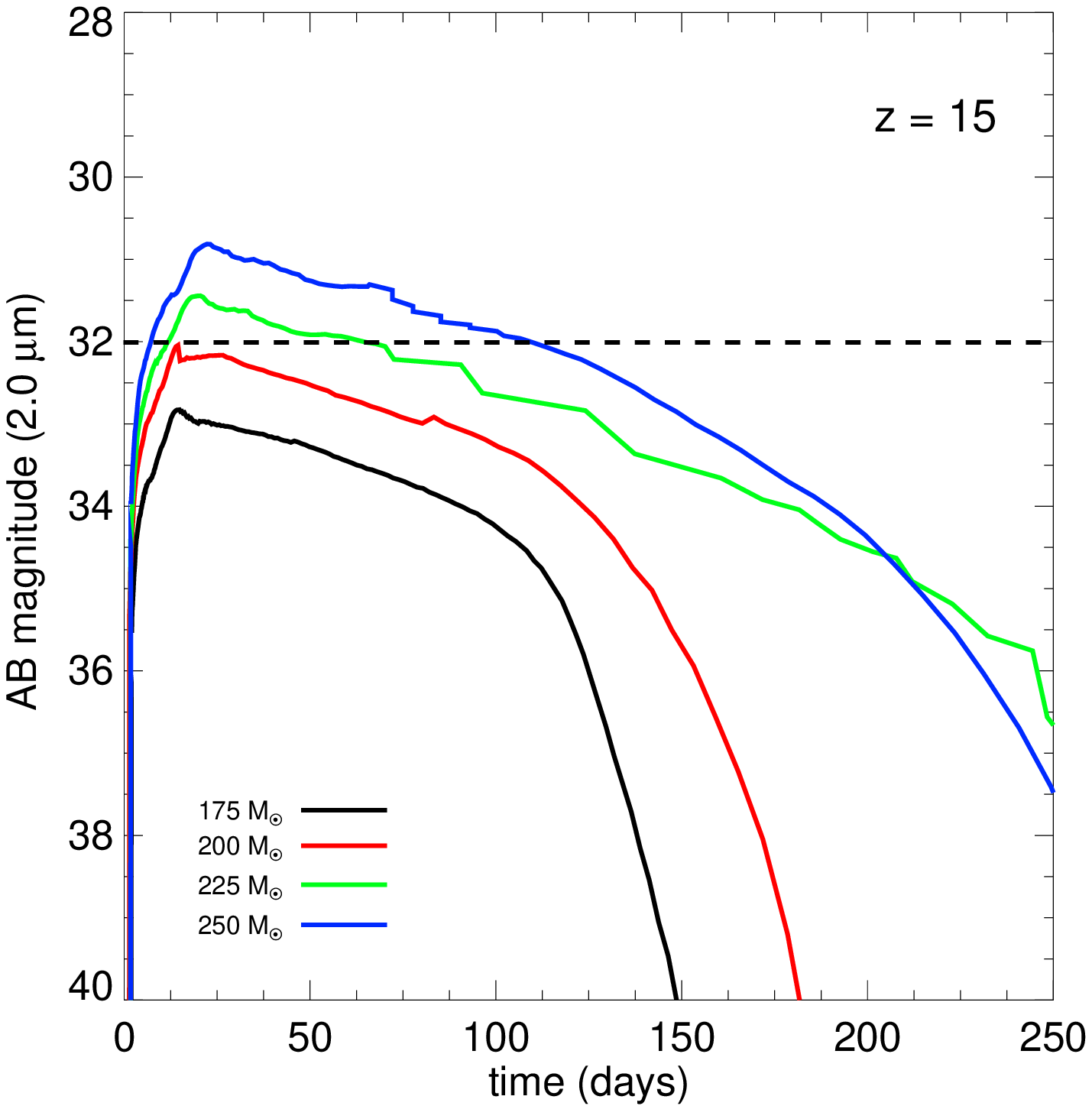,width=0.45\linewidth,clip=} 
\end{tabular}
\end{center}
\caption{NIRCam light curves for the 4 z-series PI SNe.  Left: $z = 10$.  Right:  $z = 15$.}
\label{fig:zLC}
\end{figure*}

We plot velocity and density profiles at 8.36 $\times$10$^5$ s, 7.41 $\times$ 10$^6$ s,  and 7.36 
$\times$ 10$^7$ s for the u250 explosion in the left and right panels of Fig.~\ref{fig:hydro}.  Soon
after shock breakout, and radiation has decoupled from the outer layers of the star, the velocity
profiles of the flow become essentially free expansions and are mostly self similar until the end of
the simulations at 3 yr.  The homologous expansion of the flow is also evident in the densities, 
although there is some variation in their structure deep in the ejecta over time.  This is consistent
with \citet{jw11} and \citet{chen11}, who found that any mixing in the ejecta was complete by shock
breakout.  

\section{Pop III PI SN Detection Thresholds}

We have calculated NIR light curves for our PI SNe with the synthetic photometry code described in 
\citet{su11}.  Each spectrum is redshifted to the desired value before removing the flux absorbed by 
intervening neutral hydrogen along the line of sight according to the prescription of \citet{madau95}.  
We then dim the spectrum by the required cosmological factors.  A variety of instrument filters can 
be easily accommodated by our code, which linearly interpolates the least sampled data between 
the input spectrum and filter curve to match the other.  It has additional capabilities such as 
reddening by dust that are not used here.  

\subsection{NIR Light Curves}

At $z \ga$ 7, Lyman absorption by the neutral IGM will absorb most flux at wavelengths blueward  
of 1216 \AA.  However, in principle the fireball could be brighter slightly blueward of the Lyman limit 
rather than redward even with IGM absorption \citep[see Figure 3 in][]{wet12a}.  At every redshift 
for each explosion, we calculate the NIR signal in \textit{JWST} NIRCam filters above and below the 
Lyman limit to find the filter in which the SN is brightest.  We find that for all the redshifts and PI SNe 
in our study the explosion is most luminous just redward of 1216 \AA\ in the source frame. We show 
light curves for our PI SNe at $z =$ 10, 15, 20 and 30 in Figures \ref{fig:uLC} and \ref{fig:zLC}.  At 
each redshift, light curves are plotted only for the filter in which the SN will be brightest.  Photometry 
limits for all four filters are AB magnitude 32.  All five u-series explosions are visible for 600 - 1000 
days, even at $z =$ 30.  The z-series SNe are dimmer and are only visible out to $z =$ 15 for at 
most 100 days.  They have lower bolometric and NIR luminosities and dim sooner because they 
create less \Ni\ than the u-series.  None of the u-series light curves reach the detection threshold 
before 50 days, or $\sim$ 1.5 days in the source frame at $z =$ 30, so shock breakout cannot be 
seen from Earth.  Likewise, none of the z-series SNe are visible before 10 days, or 16 hours in the
source frame.  The X-rays and hard UV in the breakout transient are absorbed by the neutral IGM.    

As noted earlier, as the fireball expands and cools its spectral peak steadily advances to longer 
wavelengths.  This is manifest in the signal in the four NIR channels in the left panel of 
Fig.~\ref{fig:LSST}.  The signal peaks at later times at longer wavelengths as the temperature of 
the fireball falls with time.  The luminosity persists for longer times at larger wavelengths because 
the shock emits at these energies for a larger fraction of its cooling time than at higher energies. 
The expansion and cooling of the explosion, together with cosmological redshifting, also accounts 
for the shift of the NIR peak to later times and longer filter wavelengths with redshift in 
Figs.~\ref{fig:uLC} and \ref{fig:zLC}.  For example, at $z =$ 10 the u-series light curves peak from 
200 - 600 days, and at $z = 30$ they peak after $\sim$ 1000 days.  As expected, the peak 
magnitude of each explosion rises with redshift.  The \textit{JWST} photometry and spectrometry
limits are AB mag 31 - 32 and 29 - 30, respectively.  Spectrometry will be possible for all five 
u-series PI SNe at $z \gtrsim$ 30 but for none of the z-series explosions at $z \gtrsim$ 10.  This 
is of note because resolving the order in which lines appear in spectra over time could provide a 
powerful probe of how heavy elements are mixed in the ejecta and thus of the explosion engine 
itself.  

As noted in \citet{wet12a}, the NIR flux evolves on timescales of $\sim$ 1000 days in the u-series
and exhibits much more variability than the bolometric flux in the observer frame (which evolves 
over 40 - 90 yr) because of the expansion and cooling of the fireball.  Such variability is the key to 
discriminating these events from primitive galaxies, with which they otherwise overlap in color-color
space.  Their NIR flux rises much more quickly than it falls, so it is easiest to detect them in their
earliest stages but they exhibit enough variation over survey times of 1 -- 5 yr to be identified at later 
stages as well.  If Pop III PI SNe are found in the NIR, they will be thousands of times brighter than 
the halo or primitive galaxy that hosts them. Indeed, if a $z \sim 15$ object exhibited any variation in 
luminosity in a survey, in all likelihood it would be a PI SN in a primeval galaxy.

\subsection{WFIRST, Euclid \& WISH}

Pop III PI SNe could be found in large numbers in future all-sky NIR surveys such as \textit{Euclid}, 
WFIRST, and the Wide-field Imaging Surveyor for High-Redshift (WISH), whose target sensitivities 
at 2 $\mu$m are AB magnitudes 24, 27, and 27, respectively.  Spectrum stacking is likely to extend 
the NIR detection limits of WFIRST and WISH to $\sim$ AB mag 29.  If so, it is clear from 
Fig.~\ref{fig:uLC} that WFIRST and WISH would detect u-series explosions out to $z \sim$ 15 - 20.  
Our calculations indicate that even at $z =$ 7, all nine PI SNe will be above magnitude 25, so 
\textit{Euclid} will only detect them below this redshift (although higher $z$ might be possible with 
spectrum stacking).  It may be that the optimal redshift range for locating Pop III PI SNe is $z \sim$ 
15 - 20 because of Lyman-Werner (LW) UV feedback.  LW backgrounds from the first stars are 
thought to destroy H$_2$ molecules and suppress cooling in primordial halos, causing them to grow 
more massive before their interiors can self-shield from LW photons, form H$_2$ and host primordial 
star formation.  Detailed numerical simulations show that the larger virial temperatures at the centers 
of such halos elevate cooling rates per H$_2$ molecule there by 2 orders of magnitude, leading to 
higher central collapse rates that favor the formation of very massive stars \citep{on07,wa08b}.  LW 
backgrounds sufficient to postpone Pop III star formation likely did not arise until $z \sim$ 20, 
delaying baryon cooling and collapse in halos until $z \sim$ 15 - 20.  Global feedback may therefore 
enhance Pop III PI SNe rates at slightly lower redshifts that are well within the range of WFIRST and 
WISH.

\subsection{LSST/Pan-STARRS}

\begin{figure*}
\begin{center}
\begin{tabular}{cc}
\epsfig{file=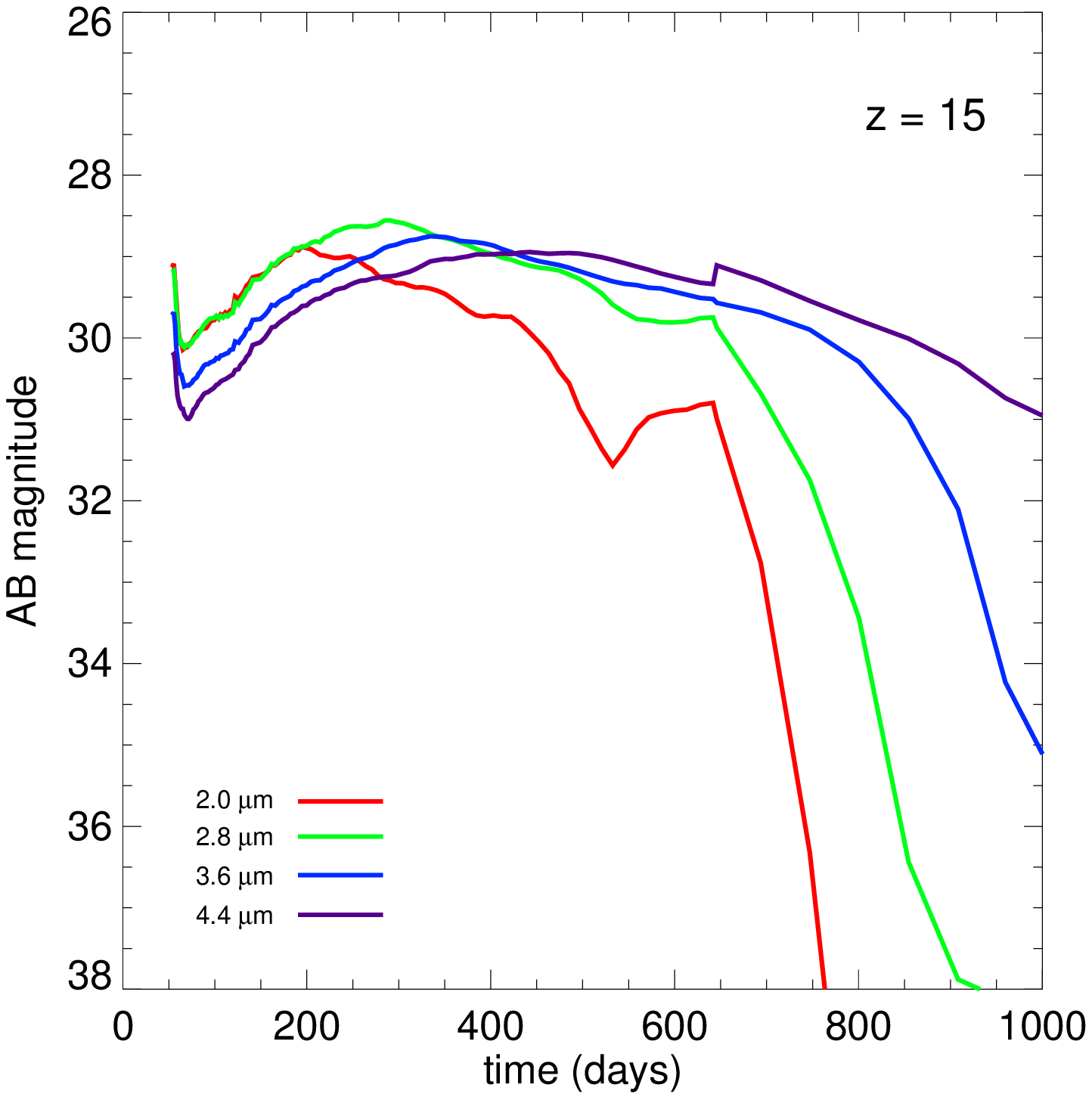,width=0.45\linewidth,clip=} & 
\epsfig{file=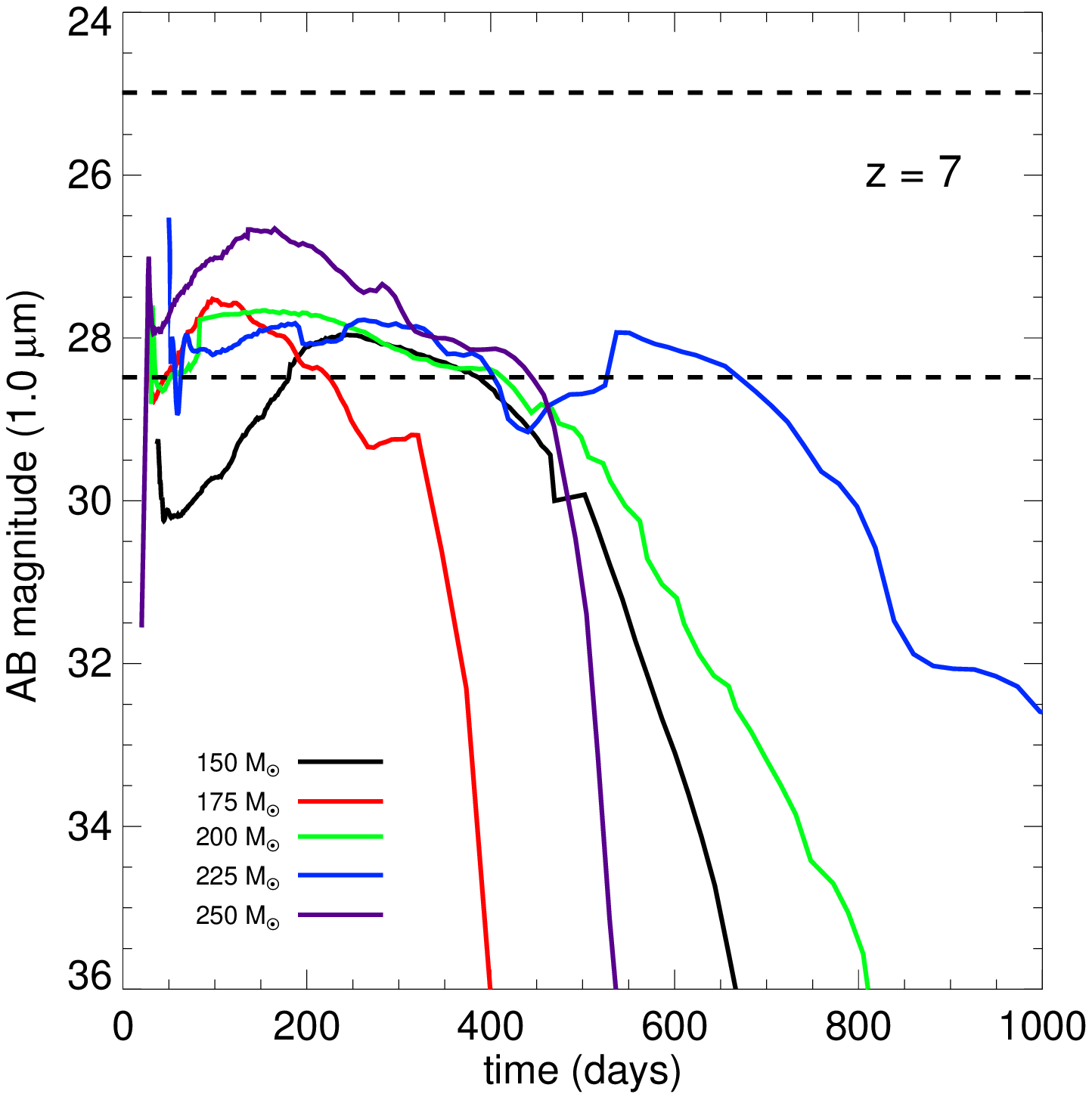,width=0.45\linewidth,clip=} 
\end{tabular}
\end{center}
\caption{Left:  spectral evolution of the u175 fireball in the NIR at $z =$ 10.  Right:  Y-band (1.0 
$\mu$m) light curves for all 5 u-series PI SNe at $z =$ 7.  The z-series explosions are too dim 
to be detected by LSST or Pan-STARRS at this redshift.  The dashed horizontal lines at AB mag 
28.5 and 25 are Pan-STARRS and LSST Y-band detection limits, respectively (these thresholds 
assume spectral stacking).}
\label{fig:LSST}
\end{figure*}

Could high-$z$ PI SNe be detected in all-sky optical and NIR surveys by the Large Synoptic Survey 
Telescope (LSST) or the Panoramic Survey Telescope \& Rapid Response System (Pan-STARRS)?  
Unfortunately, above $z \sim 10$ source frame wavelengths that would be redshifted into the optical 
are extinguished by Lyman absorption.  But LSST will have a Y-band (0.95 - 1.070 $\mu$m) limit of 
AB magnitude 22 that may be extended to 25 with spectrum stacking.  Pan-STARRS has a Y band 
detection limit of AB magnitude 26 that can be extended above 28 by spectrum stacking.  As shown 
in the right panel of Fig.~\ref{fig:LSST}, the five u-series explosions will be visible to Pan-STARRS at 
$z =$ 7 but not to LSST.  As we discuss in greater detail below, it has been speculated that very 
massive Pop III stars could form down to $z \sim$ 6 in pockets of metal-free gas; if so, their SNe 
could be detected by Pan-STARRS.

\section{Pop III PI SN Detection Rates}

Although \textit{JWST} will clearly be sensitive enough to detect $z \gtrsim$ 30 PI SNe, will it 
encounter such explosions over reasonable survey times given its narrow field of view?  Their
detection in a given survey critically depends on their event rates, which in turn are governed by 
primordial star formation rates (SFRs) and the Pop III IMF.  Many physical processes regulate 
the Pop III SFR over cosmic time.  Metals and UV feedback from early generations of stars are  
especially important, since Pop III stars can only form in pristine gas and LW photons can destroy 
the H$_2$ required for baryons in halos to cool and collapse into stars \citep[e.g.,][]{hrl97,gb01,
met01,wa07,on08}.  

\subsection{Semi-Analytical Estimates}

The original estimates of Pop III PI SN rates were based on simple halo mass distributions and 
cosmological parameters that lead to first star formation at earlier epochs than do more improved 
parameters today \citep[e.g.,][]{wmap7}.  They predict event rates that range from 0.1 - 1.5 sq. 
deg$^{-1}$ yr$^{-1}$ at $z \sim$ 25 \citep{wa05,wl05} to 0.2 and 4 sq. deg$^{-1}$ yr$^{-1}$ at $z 
\sim$ 25 and 15, respectively \citep{wl05}.  These rates exclude clustering and radiative and 
mechanical feedback between halos \citep[e.g.,][]{oet05,mjh06,mcint06,wet08b,wet10} but do 
consider global LW UV backgrounds that can delay star formation to slightly later epochs.  

\subsection{Numerical Simulations}

Cosmological simulations that incorporate chemical and radiative feedback in varying degrees of 
detail have produced more realistic Pop III SFRs for the first billion years of cosmic evolution 
\citep[e.g.,][]{tfs07,tss09,get10,maio11,hum12,jdk12,wise12}. They basically agree on event rates 
at high redshifts but differ somewhat from the original analytical estimates, mostly because they 
use more recent cosmological parameters.  By modeling the rise of LW backgrounds and taking a 
simple approach to early metal enrichment, \citet{hum12} find a cumulative rate of $\sim$ 0.5 - 5 
sq. deg$^{-1}$ yr$^{-1}$ for Pop III PI SNe at $z$ $\ga$ 5.  \citet{jdk12} find PI SN rates of $\sim$
0.3 sq. deg$^{-1}$ yr$^{-1}$ over the same redshift range, which are slightly lower because their
models employ more sophisticated prescriptions for chemical and mechanical feedback by SNe 
and assume a less top-heavy Pop III IMF.  They adopted a PI SN progenitor mass range of 140 - 
260 \Ms\ \citep[e.g.,][]{hw02}, but new models have since extended this range down to $\sim$ 65 
\Ms\ \citep{cw12}.  This new lower limit revises the PI SN rate in \citet{jdk12} upward by a factor of 
a few because of their choice of a Salpeter-like slope for the IMF.

\subsection{$z =$ 15 - 20 and $\ga$ 25}

The cumulative PI SN rates reported by \citet{hum12} imply 10$^{-3}$ - 10$^{-2}$ events per yr
per \textit{JWST} NIRCam field of view (10 arcmin$^2$).  Thus, \textit{JWST} should be able to
find 5 - 10 PI SNe over the lifetime of the mission, but greater numbers may be found if more 
time is dedicated to the surveys.  We note that Lyman absorption above $z \sim$ 6, which we 
include in our study, may slightly reduce the detection rates predicted by \citet{hum12}.  All-sky 
surveys will find far greater numbers of PI SNe, albeit at somewhat lower redshifts. Given the PI 
SN rates reported by \citet{hum12} and \citet{jdk12}, up to $\sim$ 10$^3$ PI SNe per year could 
be found at 15 $< z <$ 20 by WFIRST and WISH.  

\subsection{$z \la$ 10}

As noted earlier, PI SNe may also be found in all-sky optical and NIR surveys by LSST or 
Pan-STARRS but at lower redshifts ($z \la$ 10) because of extinction by Lyman absorption 
\citep[see also][for a recent PI SN detection campaign in \textit{Spitzer} data]{fsmk09}.  Thus, if 
very massive Pop III stars form at 6 $\la$ $z$ $\la$ 10, these surveys may discover their SNe if 
their rates are sufficiently high.  Several recent numerical simulations suggest that such stars 
could form in isolated pockets of metal-free gas at $z \la$ 6 \citep{tfs07,ts09,tss09}, and such 
environments have now been discovered at even lower redshifts \citep{fop11}.  Including LW 
backgrounds and a simple prescription for metal enrichment based on a numerical simulation, 
\citet{tss09} estimate a cumulative Pop III PI SN rate of $\sim$ 10$^{-2}$ yr$^{-1}$ deg$^{-2}$ 
at 5 $\la$ $z$ $\la$ 10, which implies an all-sky rate of up to $\sim$ 10$^3$ yr$^{-1}$.  

Other cosmological simulations that include chemical enrichment and mechanical feedback by
SNe have found lower Pop III SFRs \citep[e.g.,][]{tfs07,maio11} that imply PI SN rates that are
smaller than \citet{tss09}.  However, newer models with better treatments of LW feedback and 
chemical enrichment now suggest that PI SN rates in this redshift range are higher by roughly 
an order of magnitude than those of \citet{tfs07} and \citet{maio11} \citep{hum12,jdk12}.  Since 
the Pop III SFR down to $z$ $\sim$ 7 reported by \citet{wise12} is in good agreement with that 
found by \citet{jdk12}, their results are also consistent with a high PI SN rate.\footnote{See also 
\citet{ahn12}, who find a similar evolution of the LW background due to massive Pop III stars at 
high-$z$.}  Overall, the results of numerical simulations indicate that the all-sky Pop III PI SN 
rate at 6 $\la$ $z$ $\la$ 10 may lie between $\sim$ 10$^3$ - 10$^4$ yr$^{-1}$, which bodes 
well for their detection by LSST and Pan-STARRS.

\section{Conclusion}

We find that Pop III PI SNe will be visible in deep field surveys by \textit{JWST} out to $z \ga$ 30 
and in all-sky surveys by WFIRST out to $z \sim$ 15 - 20.  They occur at rates that are sufficient 
to appear in deep field searches but will be discovered in much greater numbers in all-sky surveys 
that can be followed up by \textit{JWST} and ground-based instruments.  It may also be possible 
to find PI SNe at lower redshifts in current surveys, for example as Lyman break dropouts in the 
\textit{Hubble Space Telescope} (\textit{HST}) CANDELS survey or with the new \textit{Subaru} 
Hyper Suprime Cam.  Such strategies will be the focus of future studies.

Our calculations emphasize the detection of Pop III PI SNe in the first few years of the explosion, 
but could their remnants be detected at later times by different means?  \citet{wet08a} found that 
PI SNe in ionized halos eventually emit up to half of the original energy of the explosion as H and 
He lines as the remnant sweeps up and shocks the relic H II region.  Unfortunately, the luminosity 
of these lines is too low and redshifted to be directly detected.  However, PI SNe can also deposit 
up to half of their energy into CMB photons by inverse Compton scattering, and could impose 
excess power on the CMB at small scales \citep{oh03,ky05,wet08a}.  The resolution of current 
ground-based CMB telescopes such as the \textit{Atacama Cosmology Telescope} and South Pole 
Telescope approaches that required to directly image Sunyaev-Zeldovich (SZ) fluctuations from 
individual Pop III PI SN remnants, so future observatories may detect them. 

The extreme NIR luminosities of primordial PI SNe could contribute to a NIR background excess, 
as has been suggested for Pop III stars themselves \citep[i.e.][]{kashl05}. X-rays from gas plowed 
up by PI SN remnants, together with radiation from early black holes, would also build up an x-ray 
background at high redshifts.  Indeed, \citet{jk11} have determined that x-rays from Pop III SN 
remnants may have accounted for $\sim$ 10\% of the reionizing photon budget at early times.  
New calculations show that PI SNe will probably not appear at 21 cm because of their lower event 
rates at high redshift and because their remnants will not emit enough synchrotron radiation to be 
directly detected by existing or future 21 cm observatories \citep{mw12}.  The imprint of Pop III PI 
SNe on the CMB and NIR backgrounds will be addressed in future studies.

If, as some numerical simulations and stellar archaeology suggest, gas in primordial halos 
fragmented into multiple Pop III stars that were tens of solar masses instead of hundreds, then 
CC SNe also occurred in the primeval universe.  Such explosions would be similar in brightness 
to those in the local universe today because their central engines mostly depend on the structure 
of the inner 3 - 4 \Ms\ of the star, which does not vary strongly with metallicity \citep{cl04,wh07,
wf12}.  Because they are 100 times dimmer than PI SNe and have softer spectra, less of their 
luminosity survives Lyman absorption by the neutral IGM at high redshift.  However, because a 
dozen or more such stars may form in the halo, CC SNe may be more plentiful than PI SNe, 
which would enhance their likelihood of detection.  Furthermore, if the star ejects a massive shell 
before exploding, the SN ejecta will light up the shell in the UV upon crashing into it 
\citep[superluminous Type IIn SNe --][]{nsmith07a,vmarle10}. Such events can have luminosities 
that can rival those of PI SNe, and they might be visible at higher redshifts than Type II explosions 
\citep{moriya10,tet12,moriya12}.  A new class of supermassive Pop III PI SNe has also now been 
discovered in numerical simulations that may be associated with the births of SMBH seeds \citep{
montero12,heg13,jet13a,wet13a,wet13b}. We have calculated light curves and spectra for all three 
kinds of explosions and found that they can be seen at $z \sim$ 15 - 20 by both \textit{JWST} and 
WFIRST\citep{wet12c,wet12e,wet12d} \citep[see also][]{tomin11}.

A few Pop III stars may die in gamma ray bursts \citep[GRBs; e.g.,][]{bl06a,wang12}, driven either 
by the collapse of very massive rapidly rotating stars \citep{suwa11,nsi12} or binary mergers with 
other 20 - 50 \Ms\ stars \citep[e.g.,][]{fw98,fwh99,zf01,pasp07}.  This is corroborated by the fact 
that a fraction of Pop III stars have been found to form in binaries in simulations \citep{turk09}. 
Although x-rays from these events can trigger \textit{Swift} or its successors such as the Joint 
Astrophysics Nascent Universe Satellite \citep[JANUS,][]{mesz10,Roming08,burrows10}, their 
afterglows \citep{wet08c} are more likely to be detected in all-sky radio surveys by the Extended 
Very-Large Array (eVLA), eMERLIN and the Square Kilometer Array (SKA) \citep{ds11} due to their 
low event rates.  We are now evaluating detection limits for Pop III GRBs in a variety of circumstellar 
environments \citep{met12a,mes13a}.  

Finally, strong gravitational lensing by massive intervening galaxies and clusters at $z \sim$ 0 - 
1 could boost flux from Pop III SNe, more than compensating for Lyman absorption and 
improving prospects for their detection \citep{rz12}.  The probability that flux from a Pop III SN 
would be lensed in an all-sky survey and the degree of magnification both depend on the event 
rate at the given redshift, and may be fairly low.  We have performed preliminary calculations 
that place the likelihood of lensing of $z \sim$ 20 objects at $\sim$ 1 - 5\% for flux boosts of 2 - 
5.  Much higher boosts (10 - 300) are possible near the edges of massive clusters but with much 
smaller search volumes and lower probabilities of encountering high-$z$ SNe.   We are now
developing and refining Markov Chain Monte Carlo ray-tracing models of strong gravitational 
lensing of $z \sim$ 20 events, the highest redshifts ever attempted, in order to assess its 
potential to reveal primeval SNe and galaxies.  Although strong lensing is not necessary for 
detecting PI SNe it may be key to finding CC SNe in protogalaxies prior to reionization, less of 
whose flux survives Lyman absorption but whose higher event rates may favor the magnification 
of this flux.

The detection of primordial SNe will directly probe the Pop III IMF for the first time and reveal the 
environments in which they form.  Their event rates will also trace the evolution of the first stellar 
populations.  Pop III PI SN explosions at $z \sim$ 10 - 15 will mark the positions of primeval 
galaxies on the sky which might not otherwise be found by \textit{JWST} or TMT.  Their discovery 
will open a direct window on the era of first light.

\acknowledgments

We thank the anonymous referee, whose comments improved the quality of this paper.  DJW 
is grateful for helpful discussions with Edo Berger, Ranga Ram Chary, Daniel Kasen, Avi Loeb, 
Pete Roming and the many participants at First Stars and Galaxies: Challenges for the Next 
Decade, held at UT Austin March 8 - 11, 2010.  He also acknowledges support from the Bruce 
and Astrid McWilliams Center for Cosmology at Carnegie Mellon University and from the 
Baden-W\"{u}rttemberg-Stiftung by contract research via the programme Internationale 
Spitzenforschung II (grant P- LS-SPII/18). JLJ was supported by a LANL LDRD Director's 
Fellowship.  MS thanks Marcia Rieke for making the NIRCam filter curves available and was 
partially supported by NASA JWST grant  NAG5-12458.  DEH was supported from the National 
Science Foundation CAREER grant PHY-1151836.  AH was supported by the US Department 
of Energy under contracts DE-FC02-01ER41176, FC02-09ER41618 (SciDAC), and 
DE-FG02-87ER40328. SEW was supported by the National Science Foundation grant 
AST-0909129 and the NASA Theory Program grant NNX09AK36G.  Work at LANL was done 
under the auspices of the National Nuclear Security Administration of the U.S. Department of 
Energy at Los Alamos National Laboratory under Contract No. DE-AC52-06NA25396. All RAGE 
and SPECTRUM calculations were performed on Institutional Computing (IC) and Yellow network 
platforms at LANL (Mustang, Pinto, Conejo, Lobo and Yellowrail).

\bibliographystyle{apj}
\bibliography{refs}

\end{document}